\documentclass[sigconf,monochrome]{acmart}

\usepackage{bbding}
\usepackage{threeparttable}
\usepackage{subfigure}
\usepackage{verbatim}
\AtBeginDocument{%
  \providecommand\BibTeX{{%
    \normalfont B\kern-0.5em{\scshape i\kern-0.25em b}\kern-0.8em\TeX}}}

 \setcopyright{none}
 
\begin{document}

\title[Learning from Home: A Mixed-Methods Analysis of Live Streaming Based Remote Education Experience\\ in Chinese Colleges during the COVID-19 Pandemic]{Learning from Home: A Mixed-Methods Analysis of Live Streaming Based Remote Education Experience in Chinese Colleges during the COVID-19 Pandemic}

\author{Zhilong Chen$^{1+}$, Hancheng Cao$^{2+}$, Yuting Deng$^{3}$, Xuan Gao$^{1}$, Jinghua Piao$^{1}$, Fengli Xu$^{1}$, Yu Zhang$^{4}$, Yong Li$^{1}$}
\authornote{Corresponding author.\\$^{**}$Equal contribution.}
\affiliation{%
  \institution{$^{1}$ Beijing National Research Center for Information Science and Technology (BNRist),\\
Department of Electronic Engineering, Tsinghua University}
  }
\affiliation{%
  \institution{$^{2}$ Department of Computer Science, Stanford University $^{3}$ The College, The University of Chicago}
  }
\affiliation{%
  \institution{$^{4}$ Institute of Education, Tsinghua University}
  }
\email{liyong07@tsinghua.edu.cn}
\renewcommand{\shortauthors}{Chen and Cao, et al.}
\newcommand{\czl}[1]{{\textcolor{red}{ [czl: #1]}}}
\newcommand{\hancheng}[1]{{\textcolor{red}{ [hancheng: #1]}}}
\newcommand{\dyt}[1]{{\textcolor{red}{ [dyt: #1]}}}
\begin{abstract}
The COVID-19 global pandemic and resulted lockdown policies have forced education in nearly every country to switch from a traditional co-located paradigm to a pure online “distance learning from home” paradigm. Lying in the center of this learning paradigm shift is the emergence and wide adoption of distance communication tools and live streaming platforms for education. Here, we present a mixed-methods study on live streaming based education experience during the COVID-19 pandemic. We focus our analysis on Chinese higher education, carried out semi-structured interviews on 30 students, and 7 instructors from diverse colleges and disciplines, meanwhile launched a large-scale survey covering 6291 students and 1160 instructors in one leading Chinese university. Our study not only reveals important design guidelines and insights to better support current remote learning experience during the pandemic, but also provides valuable implications towards constructing future collaborative education supporting systems and experience after pandemic.
\end{abstract}

\ccsdesc[500]{Human-centered computing~Empirical studies in HCI}
\ccsdesc[500]{Human-centered computing~Empirical studies in collaborative and social computing}
\ccsdesc[500]{Applied computing~Distance learning}
\ccsdesc[500]{Applied computing~E-learning}

\keywords{LS learning, live streaming, distance learning, COVID-19}


\maketitle

\section{Introduction}
The COVID-19 global pandemic and resulted lockdown/social distancing policies have vastly transformed education, bringing the traditional co-located paradigm to a pure online distance learning paradigm, which generates new education models~\cite{yang2020turn}. As billions of students leave their school and proceed with their studies from home, instructors turn to various technologies to ensure the sustainability of learning~\cite{liu2020ensuring}. Lying in the center of this learning paradigm shift is the wide adoption of live streaming based learning (LS learning) mode, where teachers connect with students through distance communication tool (e.g. Zoom, live streaming platforms), give lectures, hold sessions and interact with students via software interfaces in real-time. Meanwhile, students rely on these platforms to collaborate and group learn with peers. 

\begin{figure} [t]
\centering
\subfigure{\includegraphics[width=.32\textwidth]{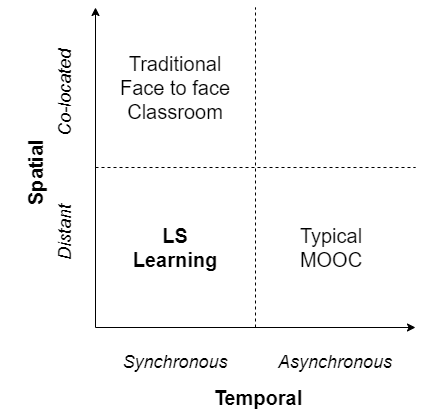}}
\caption{Various education formats characterized by temporal and spatial dimensions.} 
\label{2X2}
\end{figure}

As demonstrated in Figure~\ref{2X2}, live streaming based education deviates from traditional learning experiences in many important ways: traditional classroom corresponds to a co-located synchronous interaction mode where all teacher-student and student-student interaction takes place live at the same location~\cite{shannon2017better}. The typical massive online open course (MOOC)~\cite{zheng2015understanding}
belongs to a distant asynchronous interaction mode, where prerecorded videos are distributed online and learners {\color{blue}asynchronously interact with instructors and} could access the content any time they want. In contrast, LS learning belongs to a distant synchronous mode, where interaction happens in real-time yet physically far apart. Even though similar experiences as LS learning have emerged before the pandemic, e.g.~\cite{chen2019integrating}, they are limited in coverage and restricted to very specific learning purposes. In comparison, current LS learning is distinct in that it is 1) large-scale in coverage, 2) ubiquitous and widely used across disciplines, and 3) primarily aimed for long-term formal learning process rather than informal experience, e.g., informal language learning~\cite{chen2019integrating}. {\color{blue}We seek to uncover how LS learning experiences differ from more traditional educational formats.}

Given the importance of LS learning under the ongoing COVID-19 pandemic, as well as its expected significant roles in the upcoming hybrid educational era, it is important to study how better LS learning experiences can be supported. We present this paper to fill in this research gap. Specifically, we ask: how has LS learning support student learning from home experience during the pandemic? What are the issues and challenges under the current setups and how may we address them through future design? Perhaps most interestingly, are there any human-computer interaction (HCI) lessons we can take away from this unprecedented natural COVID-19 ``learning from home'' experiment that can guide us in designing future education tools after the pandemic? 

To shed light on these questions, we present a {\color{blue}mixed-methods} study on LS learning experiences during the COVID-19 pandemic. With a special focus on Chinese higher education, we conducted semi-structured interviews on 30 students, and 7 instructors from diverse disciplines across different universities, meanwhile administered a \emph{large-scale} survey covering 6291 students and 1160 instructors in one of the leading Chinese universities. We focus our analysis on 1) individual learning experience, and 2) interaction experience between instructor-student and student-student. Our findings suggest live streaming based education do help students and teachers achieve their education goal to a great extent, yet there are several key challenges emerging under the current paradigm, including students' difficulties in paying continuous attention, decreased learning efficacy, and lack of engagement/collaboration experiences. We further demonstrate how various interaction formats, including audio, video, text box, danmaku, quiz, vote, in the LS learning platforms enable novel learning experiences, which contributes to variations in instructor-student and student-student relationships. Based on our analysis, we further propose and discuss several teaching and learning practices for courses of different characteristics. Finally, we propose important design guidelines and insights to better support LS learning experiences during the pandemic, and offer valuable implications towards constructing future collaborative education supporting systems and experiences post-pandemic.

Contributions of this paper can be summarized as follows: 
\begin{itemize}
\item{We present a large-scale systematic analysis of LS learning experiences through mixed methods in Chinese colleges.}

\item{We reveal challenges and possibilities with regard to LS learning, and provide concrete guidelines on how to enable better LS learning experience given the current technology.}

\item{Our study points to several important design implications on future educational tools to support LS learning, e.g., balance of anonymity and real-name systems.}

\end{itemize}
\section{Related Work and Background}
{\color{blue}We first position our work in the rich literature from the following aspects: formats of education (the activity we study), remote collaboration tools/live streaming platforms (the technology our study focuses on), and the adoption of such technology in activity (agency), where we present an overview of popular LS learning platforms used by Chinese colleges during the COVID-19 pandemic.}

\subsection{Education Formats}
Past human-computer interaction (HCI) and computer-supported cooperative work (CSCW) researches have shown that spatial distance~\cite{olson2000distance} and temporal synchronization~\cite{hrastinski2008asynchronous} would exert immense distinctions in experiences in the education domain. With these two dimensions as classification criteria, education can be divided into 4 categories: co-located and synchronous, distant and synchronous, co-located and asynchronous (rarely implemented), and distant and asynchronous (see Figure~\ref{2X2}). 

Traditional classes typically belong to the co-located and synchronous format. Past researchers attempted to develop tools for digitizing and supplementing face-to-face learning, helping provide peer feedback~\cite{shannon2017better}, support reflection, communication and planning~\cite{kharrufa2017group}, deliver quizzes~\cite{poon2019engaging}, and reflect on teachers' performances~\cite{an2019unobtrusively}, etc. {\color{blue}In terms of the distant and asynchronous education format, a representative exemplification is the typical Massive Open Online Courses (MOOCs) format: interactions between instructors and students are largely asynchronous and distant, \emph{i.e.}, most MOOC education uses prerecorded lecture videos, while recent studies have taken steps to introduce synchronicity to MOOCs such as MOOC chatrooms~\cite{coetzee2014chatrooms,kulkarni2015talkabout}.} Research efforts on MOOCs have been dedicated to not only investigating students'~\cite{zheng2015understanding} and instructors'~\cite{zheng2016ask} overall motivations and perceptions towards MOOC usages, but also understanding and modeling specific MOOC features, including forum use~\cite{coetzee2014should}, virtual team formation~\cite{wen2017supporting}, geographic diversity in MOOC discussions~\cite{kulkarni2015talkabout}, divided attention~\cite{xiao2017undertanding}, and MOOCs' support on employability~\cite{dillahunt2016massive}. Some other research on distance education focused on paid degree programs. For instance, Sun, Rosson, and Carroll~\cite{sun2018community} investigated community among online learners in remote learning programs. Sun, Wang, and Rosson~\cite{sun2019distance} uncovered how distance learners connect with a special focus on shared identity, focused work, and future possibilities.

However, very limited attention has been paid to the distant and synchronous learning format. Work on telepresence, such as Newhart and Olson's~\cite{newhart2017my}, has taken the preliminary steps into understanding remote learning engagement, but only from instructors' perspectives. Chen, Freeman, and Balakrishnan~\cite{chen2019integrating} revealed how different modalities shape language-learning live streaming. However, existing work on synchronous distant learning is constrained to voluntary/informal learning, the context and motivations of which is far different from formal education, e.g., colleges. Different from them, in this work we examine the live streaming based remote education experience -- the form of distant and synchronous learning enabled and forced by COVID-19 at scale in Chinese colleges, where studies are carried out under formal education settings. 

\subsection{Remote Collaboration Tools and Experience}
Remote collaboration has long been a central topic in CSCW literature. Much work has been done in this space to understand and support better remote collaboration tools and experience \cite{olson2000distance,koehne2012remote}. Two important lines of studies that closely relate to our work are performance and peer to peer relationship research under remote collaborations. With regard to performance, Gumienny et al. found that idea generation and feedback collection can be facilitated if a remote collaboration system offers real-time synchronous editing as well as asynchronous inputs~\cite{gumienny2013supporting}. A series of research has been done to understand the role of spatial audio and video in supporting more engaging remote collaboration experiences~\cite{mehrotra2011realistic, hauber2006spatiality}. Junuzovic et al.~\cite{junuzovic2011towards} studied the layout guideline for designing more effective multi-party, gaze-aware desktop videoconferencing tools. Cao et al. \cite{cao2021large} launched a large scale analysis of remote meeting multitasking behavior and studied its impact on productivity and worker well being under remote collaboration. In terms of social relationships under remote collaboration, research has been done to investigate remote team viability~\cite{cao2020my, whiting2019did}. Macaranas et al.~\cite{macaranas2013sharing} studied how watching video programs together at a distance affects team cohesion. Other works investigated the role of novel technology in shaping unique remote collaboration experience~\cite{cao2020you, cao2020your,chen2020understanding}. Building on top of these works, here we investigate how LS learning as an emerging instance of remote collaboration between student and teacher, impacts learning experience from both individual learning outcome and group collaboration effectiveness/inclusiveness perspectives.

\subsection{Live Streaming in HCI and CSCW}
As an increasingly popular medium, live streaming has attracted the attention of numerous researchers in the HCI community. One line of work focused on the general usage of live streaming and highlighted its similarities and differences compared to other mediums. For instance, Juhlin, Engström, and Reponen~\cite{juhlin2010mobile} demonstrated what contents are shared on these platforms and how people manage these contents. Dougherty~\cite{dougherty2011live} evaluated live streaming from the civic engagement angle. Tang, Venolia, and Inkpen~\cite{tang2016meerkat} characterized motivations behind live streaming, and Lu et al.~\cite{lu2018you} focused on Chinese practices as a case study. Haimson and Tang~\cite{haimson2017makes} identified immersion, immediacy, interaction, and sociality as drivers of engaging live streaming experiences. Other studies focused on the use of live streaming in specific domains, for example, video games~\cite{hamilton2014streaming,lessel2017expanding}, visual art~\cite{yang2020snapstream}, intangible cultural heritage~\cite{lu2019feel}, knowledge sharing~\cite{lu2018streamwiki} and outdoor activities~\cite{lu2019vicariously}. Some recent studies have taken the first steps to leverage live streaming for education. Faas et al.~\cite{faas2018watch} examined how live streaming enables programming mentoring. Chen, Freeman, and Balakrishnan~\cite{chen2019integrating} investigated how diverse modalities can be used for live streaming to support language learning. Sun et al.~\cite{sun2019presenters} revealed how live streaming can be adopted for online lectures with audience flow
prediction. Extending these studies, we focus on understanding student experience under recent emerging live streaming enabled formal education (LS learning) during the pandemic - an important yet not well-studied application instance of live streaming.

\subsection{Live Streaming based Remote Learning Platforms in China}
In China, several platforms have been adopted for live streaming based learning \footnote{We refer to live streaming based learning as LS learning platforms for short in later analysis.} during the pandemic. These platforms share the functions of enabling synchronous and live sharing of visual and audio contents. 
Table~\ref{tab:Platform} we introduce and summarize characteristics of three major categories of LS learning platforms in China. 

\begin{table*}[!htbp]
\centering
\caption{Major categories of LS learning platforms and their characteristics.}
\begin{tabular}{cccccccc}
\toprule
\textbf{Platform} & \textbf{Example} & \textbf{Video} & \textbf{Audio} & \textbf{Textbox} & \textbf{Danmaku~\cite{lu2019vicariously}} & \textbf{Quiz} & \textbf{Vote}\\
\hline
E-Classroom & Rain Classroom & \checkmark (1)*
& \checkmark (1) & \XSolid & \checkmark & \checkmark & \checkmark\\
Video Conference & Zoom, Tencent Meeting & \checkmark (2) & \checkmark (2) & \checkmark & \XSolid & \XSolid & \XSolid\\
Traditional Live streaming & Bilibili, Kuaishou & \checkmark (1) & \checkmark (1) & \checkmark & \checkmark & \XSolid & \XSolid\\
\bottomrule
\end{tabular}
\begin{tablenotes}
\item *The number in the bracket indicates the interaction directions available (one-way/mutual interactions).
\end{tablenotes}
\setlength{\belowcaptionskip}{10pt}%

\label{tab:Platform}
\vspace*{-4mm}
\end{table*}

\begin{figure*} [t]
\centering
\subfigure[LS Learning through E-classroom: danmakus are marked in red circles, instructors' video in blue rectangles and instructors' screen sharing shown as background. ] {
\label{fig:e-class}
\includegraphics[width=.30\textwidth]{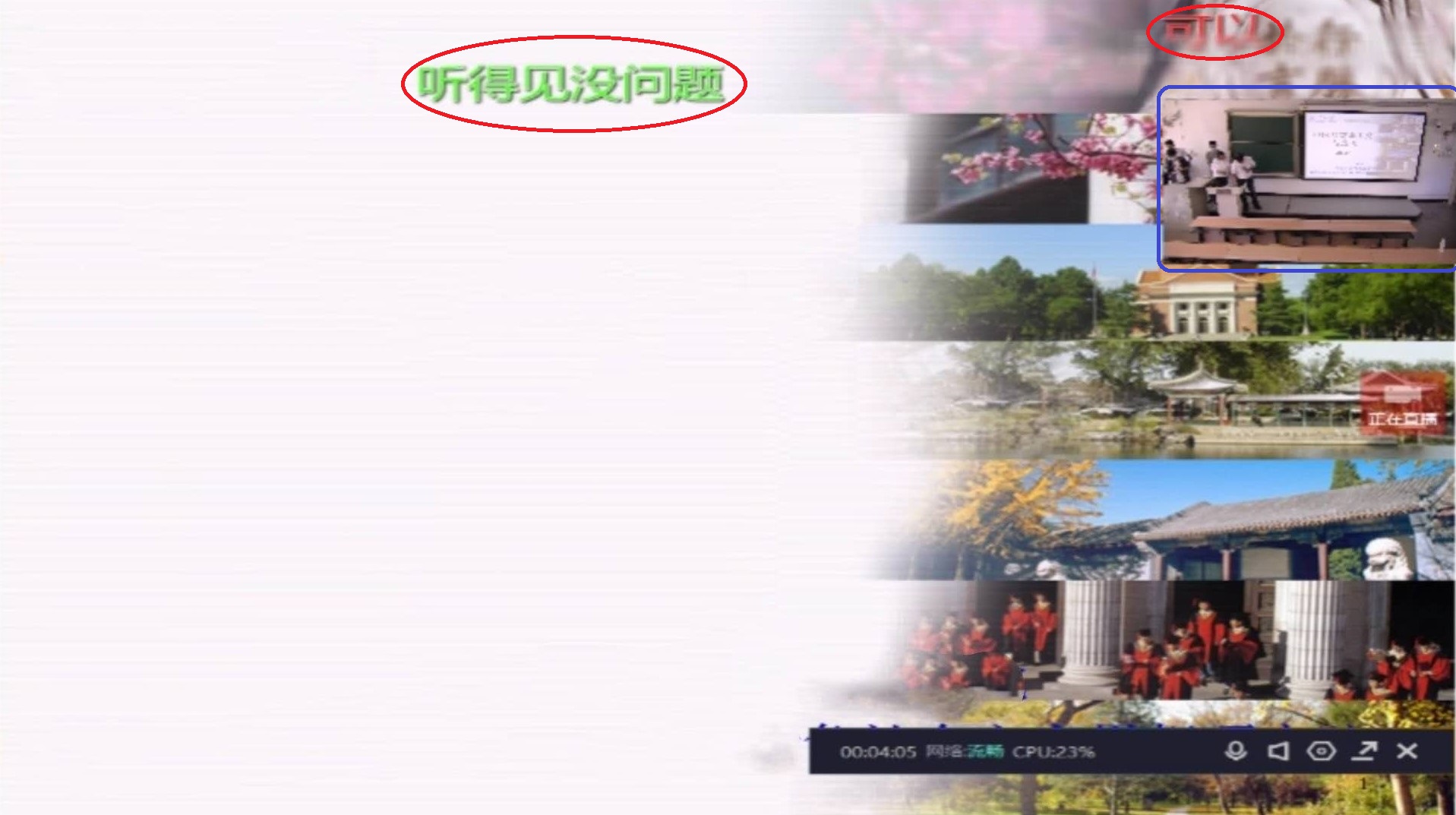}}
\subfigure[LS Learning through Video Conference: the instructors' video is marked in red rectangle and instructors' screen sharing shown as background.] {
\label{fig:video}
\includegraphics[width=.36\textwidth]{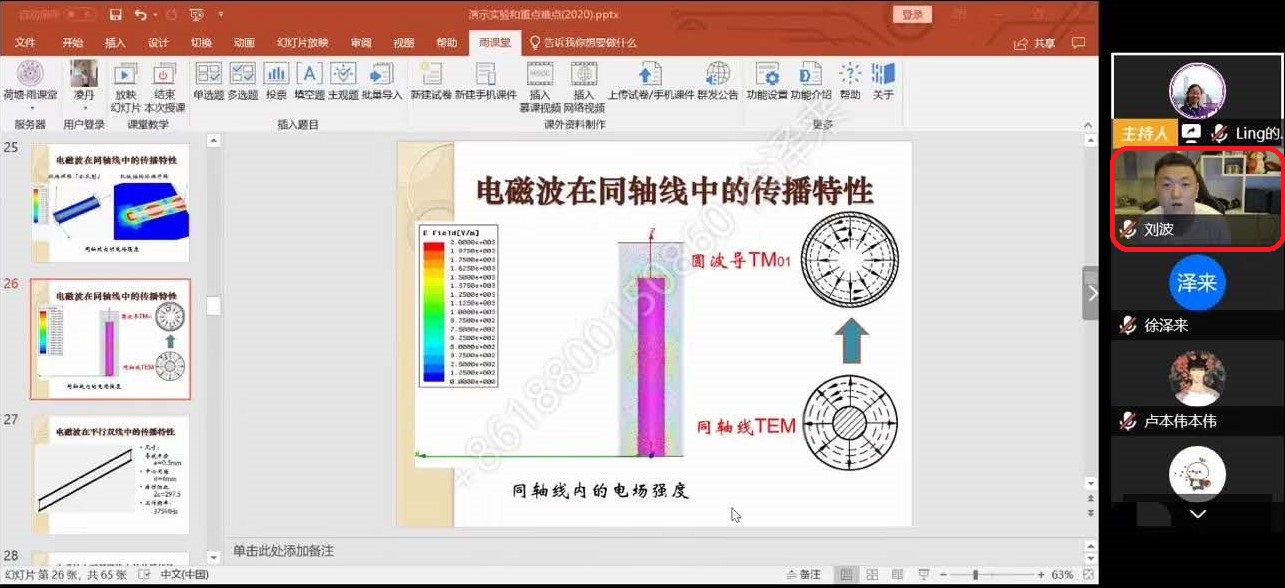}}
\subfigure[Quiz/Vote function integrated in PPT: the quiz function records who answers what and has a correct answer set in advance, while the vote function only presents the overall voting distribution.] {
\label{fig:quiz}
\includegraphics[width=.30\textwidth]{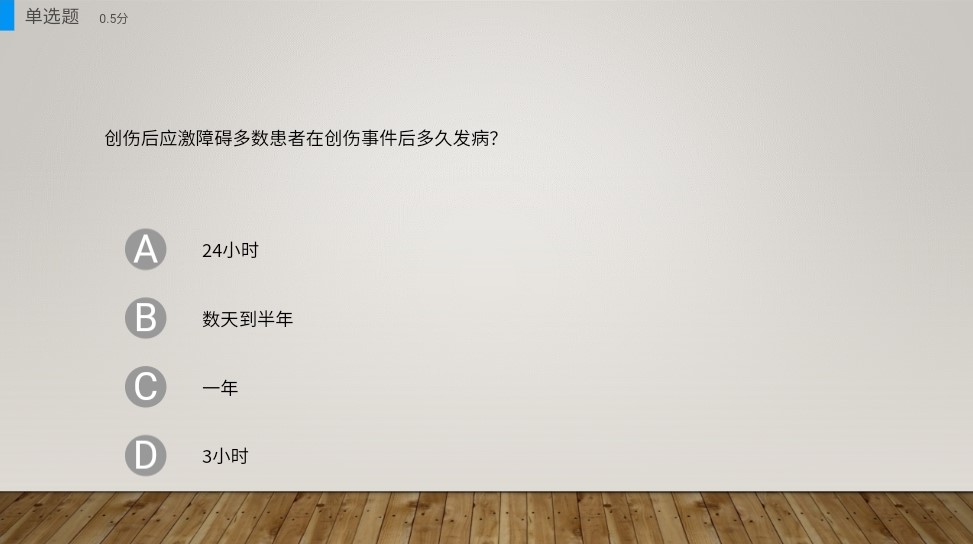}}

\caption{Demonstration of LS Learning Platforms.} \label{fig:Platform}
\end{figure*}

\textbf{\textit{E-classroom.}} A few platforms are specifically designed for distance education, which are also known as e-classrooms. These platforms support synchronized visual and audio interactions and thus enable LS learning. One widely-adopted instance is Rain Classroom, which has been utilized by a wide range of universities. In Rain Classroom, instructors can not only broadcast live video and audio and share their screens, but also launch quizzes and votes to students in the e-classroom. However, at the time we collect our data, students are not able to turn on their own cameras or microphones to share their own visual or audio information. Thus most of the time the interaction is one-way, \emph{i.e.}, from instructors to students. Nevertheless, students could interact with the instructors through textual danmaku~\cite{lu2019vicariously}, which the instructors could choose to have it turned on or off (see Figure~\ref{fig:e-class}).

\textbf{\textit{Video conference.}} As shown in Figure~\ref{fig:video}, online video conferencing platforms have also been adopted by a great many instructors~\cite{hopper2014bringing,hussa1996distance,lawson2010images} to fulfill LS learning, among which Zoom and Tencent Meeting are two most frequently used software. On these platforms, should the instructors/administrators permit, anyone in the class can turn on his or her microphone or camera to share audio or video contents. Text boxes are integrated into video conferencing tools and enable textual interactions.

\textbf{\textit{Traditional live streaming.}} Some instructors implement LS learning through live streaming platforms such as Bilibili and Kuaishou. On these platforms, instructors usually take the role of streamers and share audios, images, and screens in real-time. Students act as viewers of the live streams and interact with instructors through text boxes. Some platforms such as Bilibili also support danmaku, where textual comments in the text box are shown upon videos in a danmaku manner.

In later analysis, we refer to all of the aforementioned instances as LS learning platforms. We focus on deriving conclusions that generalize to all LS learning experiences, rather than those specific to a single platform.
\section{Method}
{\color{blue}Based on related work and background on education (activity) and remote collaboration tools/live streaming platforms (technology), in this study, we seek to gain a better understanding of user experience when live streaming are adopted at large scale in education (agency). Specifically, we ask the following research questions: 

\textbf{RQ1}: What features do instructors use under LS learning and how are they used?

\textbf{RQ2}: How does LS education influence individual learning experience?

\textbf{RQ3}: How does LS education shape collaborative learning experience and influence social relationships?}

To answer them, we adopt a mixed-methods methodology combining in-depth interviews and a large-scale survey.

\subsection{Semi-structured Interview Study}
We interviewed 30 students and 7 instructors {\color{blue}from various Chinese universities} who engaged in LS learning in Spring 2020 semester, where we tried to diversify the disciplines of the instructors and majors of students as much as possible to attain variation~\cite{corbin2014basics}. Table~\ref{tab:Student} and Table~\ref{tab:Instructor} demonstrate the detailed information of the students and instructors participated, respectively. Interviewees starting with S (S1-S30) represent students, and participants starting with T (T1-T7) are instructors. The interviews were completed either in person or through remote audio calls and took the form of a semi-structured manner. In the interviews, we penetrated into how courses were taken through LS learning during the COVID-19 crisis, how the teaching/learning experiences were and how they differed from traditional face to face learning and MOOC learning, how the learning outcomes were, how instructors and students collaborated, and how they felt their relationships. The interviews were all conducted in Mandarin, each of which lasted around 30-45 minutes and compensated with an honorarium of 50 CHN. After receiving the oral consent of the participants, we audio-taped the interviews and transcribed them leveraging transcription service and manual modifications, where we removed the identifiable information to guarantee better protection of interviewees' privacy. 

\begin{table*}[!htbp]
\centering
\caption{Summary of the basic information of interviewed students.}
\begin{tabular}{ccc|ccc|ccc}
\toprule
\textbf{ID} & \textbf{Gender} &  \textbf{Major} & \textbf{ID} & \textbf{Gender} & \textbf{Major} & \textbf{ID} & \textbf{Gender} & \textbf{Major}\\
\hline
S1 & F & Human Resources & S2 & F & Calligraphy & S3 & M & Chemistry\\
S4 & M & Electronic Engineering & S5 & F & Materials Science & S6 & M & Electronic Engineering  \\
S7 & M & Math & S8 & M & Humanity & S9 & F & Law \\
S10 & F & Medicine & S11 & M & Information Science & S12 & M & Information Science \\
S13 & F & Public Policy & S14 & F & Marine Technology & S15 & M & Information Science \\
S16 & F & Geophysics & S17 & F & Electronic Engineering & S18 & F & Law \\
S19 & F & Automation & S20 & F & Building Environment & S21 & M & Information Science \\
S22 & M & Computer Science & S23 & M & Oral Interpretation & S24 & F & Economics \\
S25 & F & Information Science & S26 & M & Information Science & S27 & F & Information Science \\
S28 & M & Mechanical Engineering & S29 & F & Art & S30 & F & Foreign Language \\
\bottomrule
\end{tabular}
\setlength{\belowcaptionskip}{10pt}%

\label{tab:Student}
\vspace*{-4mm}
\end{table*}

\begin{table*}[!htbp]
\centering
\caption{Summary of the basic information of interviewed instructors.}
\begin{tabular}{ccc|ccc}
\toprule
\textbf{ID} & \textbf{Gender} & \textbf{Discipline} & \textbf{ID} & \textbf{Gender} & \textbf{Discipline}\\
\hline
T1 & F & Foreign Languages and Literature & T2 & F & Education \\
T3 & M & Data Science & T4 & M & Telecommunication \\
T5 & F & Physical Electronics Experiment  & T6 & M & Nuclear Science and Technology \\
T7 & M & Foreign Languages and Literature \\
\bottomrule
\end{tabular}
\setlength{\belowcaptionskip}{10pt}%

\label{tab:Instructor}
\vspace*{-4mm}
\end{table*}

To analyze the interviews, we first open-coded~\cite{corbin2014basics} the transcriptions. Two Mandarin-speaking authors separately coded 5 interview transactions and appointed a discussion on the codes until reaching consensus. Then, one of these authors coded the remaining transcriptions and periodically discussed with the other author to guarantee agreements on the codes. One other native Chinese author was responsible for the translation of the codes and corresponding quotes into English and the aforementioned two authors were responsible for validating and refining the translations. Upon finishing these procedures, the whole research team thoroughly discussed the contents that had been extracted. With sub-categorization and constant comparison, we developed and continually amended the emerging themes.

\subsection{Survey Study}
To ensure the generalizability of our findings, we further launched a large-scale survey study in a leading Chinese university where several live streaming platforms and distance communication tools are officially provided for teachers to enable LS learning. The survey was distributed through a college-level administrative approach online towards (1) instructors who have taught at least one course in both Fall 2019 semester and Spring 2020 semester, and (2) students who have taken coursework in both Fall 2019 semester and Spring 2020 semester. We compared the learning experience in Fall 2019 semester (all education in traditional co-located classrooms) vs. Spring 2020 semester (all education through LS learning). We focused on the differences between the two consequent semesters so as to measure the changes brought by LS learning. Questions include learning/teaching experience, quality, outcomes, practices, and effectiveness, with a special focus on education experiences and learning and teaching behaviors. A total of 1160 instructors and 6291 students participated in and validly responded to the survey. 

The instruments utilized in the survey are adapted from previous research~\cite{riggio1993social,moeller2019loneliness}. A four-point rating scale was used with descriptions such as [`not at all', `very little', `quite a bit', `very much'] or [`never', `rarely', `sometimes', `often']. It is worth noting that due to the large-scale nature of the survey study, statistical significance testing is not appropriate as large scale data tend to signify most differences as significant. Therefore, we adopt an `effect size’ perspective instead, where a difference no less than 1/3 of a standard deviation in the mean score is considered as a meaningful effect~\cite{cohen2013statistical}.
\section{Findings}
In this section, we report the findings on LS learning experience. The major themes emerging from our mixed-methods study can be characterized into: 1) interaction format experience with specific platform features, 2) overall individual learning experience, and 3) interactive learning experience between instructor-student, student-student, and community perception.

\subsection{Interaction Formats in LS Learning}
LS learning supports a wide range of interaction formats, ranging from audio, videos, text box messages, danmakus, quizzes to votes. Each of these formats was recognized as contributing to the overall functionality of live streaming-based remote education by our participants, in which participants illustrated how these features shape live streaming-based remote education.

\textbf{\textit{Audio.}}
Audio was the most used modality by instructors. In most LS classes, only the instructor leaves the microphone on throughout the class, and students only turn it on when answering questions -- or else the class would be too noisy (T2, T3 \& T4). There were also some classes that required every student to leave their microphones on, especially small classes. This allowed students to interrupt the class at any time if they had questions (T5). It also allowed for classes that required a substantial amount of interactions between the instructor and the students. One example was during a Russian class the instructor asked everyone to turn on their microphones since \emph{“the class needed synchronous Russian practice between students, or between students and me [the instructor]”} (T1). T1 also emphasized that only small classes could allow students to leave their microphones on throughout the class: \emph{“if this was a big class, we could not do it since there would be too much background noise which would interrupt the class”}. 

Most instructors would set up separate question and answer (Q\&A) sessions to answer students’ questions after class, and in this case students would be allowed to leave their microphones on throughout the session. \emph{“Audio makes communication more convenient and accurate”}, according to S25, \emph{“thus I am more willing to ask my questions by speaking up instead of typing up”}.  

There were also some complaints about using audio. S5 raised the concern that the interaction between the instructor and the students through microphones might interrupt the pace of the class. For example, during one of his fast-paced classes, \emph{“occasionally there would be some students asking trivial questions that would slow down the pace of the whole class”} (S5). S5 also mentioned that he felt \emph{“the quality of communication online through microphones is worse than the quality of face-to-face office hours sessions in offices, due to the absence of boards and pens: it would make much more sense if the instructor could draw some graphs to explain how things work instead of pure talking”} (S5). Furthermore, S27 reported a technical issue in that one time her microphone was automatically turned on in one of her classes, and it made her feel \emph{“very awkward”} since she had been singing.

\textbf{\textit{Video.}}
Apart from audio, video was the second most frequently used modality by instructors. There were mainly two kinds of video displays combined in LS classes: 1) turning on video and displaying one’s face through the front camera and, 2) sharing one’s screen. For the majority of the classes, only the instructor turned on the camera to let students see his/her face throughout the sessions. In most live streaming sessions, the instructor would also share his/her screen so students could see lecture slides or any other materials related to the class. While sharing their screen with students, most instructors insisted on showing their faces to students. As commented by T2, \emph{“No matter how advanced the technology is, we think humanity plays an irreplaceable role in education ... it is very important to let students see us, so that they feel like they are in a real classroom”} (T2). 

In most LS classes, students would not keep their video on throughout the entire session. If everyone turned on his/her camera throughout the session, this could destabilize the Internet connection and the platform, thus diminishing class quality. Due to technical issues, T1, T3, T4, and T5 chose to turn off students’ cameras. However, many teachers emphasized the importance of seeing students’ faces in the interviews. For example, T1 would sometimes ask students to turn on their cameras for several minutes during the class to have a better sense of \emph{“students’ states”}, and T2 felt it was necessary to see every student’s face to \emph{“promote emotional interactions”}. 

From students’ perspectives, {\color{blue}according to the interview,} the majority of the students agreed that turning on their cameras \emph{“made them feel more motivated and concentrated”} during the class (e.g., S21, S23, S27 \& S29). S9 and S29 also mentioned that they liked to see everyone’s face since \emph{“it just feels like a normal class: we are sitting in a classroom and listening to the teacher”} (S9). Some students reported that a benefit of the video modality is that students could \emph{“see the instructor’s face and blackboard clearer and listen to the instructor’s voice clearer”}, while pre-pandemic only the first several rows of students in a classroom could see and hear the instructor clearly (S1 \& S12).

However, some other students, e.g., S23 \& S25, disliked video when having to show their faces to everyone: \emph{“it’s so awkward turning on the camera and I really don’t like others seeing me wearing pajamas in my bedroom”} (S23). Another concern was raised by S2, who complained about the inconvenience of using the video modality when she was practicing calligraphy in class: \emph{“previously, since the instructor would be in the classroom, the instructor could provide immediate feedback to each student if he/she found something wrong; now we have to show our calligraphy with the front camera after we are finished. Thus, we cannot receive feedback in time”} (S2).

\textbf{\textit{Text box.}}
Text box was used frequently throughout the live stream sessions by students. Students would use the text box for various purposes such as greetings (S12 \& S20), asking questions (S5, S20 \& S23), discussing problems (S6, S25 \& S27), providing feedback to instructors (S10 \& S21), etc. \emph{“Because the comments were not anonymous, we were cautious when sending comments”}, said S25. The majority of the {\color{blue}interviewees} made similar observations to S25's. Many of them also reported that \emph{“discussing problems in the text box made us closer”} (S12), pointing out that thinking about other peers’ questions was inspirational, and assisted with their own learning. 

On the other hand, teachers attended to the text box less frequently than students {\color{blue}according to the interview}. For example, both T1 and T4 admitted that they rarely checked the text box since they \emph{“forgot to check it while teaching”} (T4). Other instructor interviewees said they primarily used the text box to post simple quizzes (T2), collect feedback from students (T2), and answer students’ questions (T3 \& T5). 
In addition, the text box does not appear on the screen unless a user clicks on it. During a live streaming class, the instructor would check the text box at a specific time -- normally toward the end of the class, since \emph{“checking the text box too frequently would interrupt the pace of the class”} (S10). Thus, according to S6, \emph{“instructors rarely answer questions immediately, creating some gaps between students’ discussion and the instructor”}.

\textbf{\textit{Danmaku.}}
Besides the text box, danmaku was another modality used frequently by the students throughout the LS classes. Danmaku refers to a form of user comments that are displayed over videos~\cite{ma2017video}. In the scenario of live streaming, danmakus are real-time user comments that are displayed over the live video~\cite{lu2019vicariously}. During LS classes, students can send danmaku at any time, and it is displayed in real-time at the top of the screen, visible to everyone in the class, including the instructor. Danmaku would appear at the top right of the screen, and move from right to left, and eventually disappear. Unlike the text box, danmaku was anonymous, which made the comments \emph{“less formal and much more casual”} (S12), and thus \emph{“encouraging more students, especially those who were too shy to speak up in front of people and to participate in the class”} and \emph{“building a stronger tie between the instructor and the students”} (S8). The majority of the students from the interview provided comments similar to S8’s, including S4, S6, S12, S16, S18, S20, S27, etc. 

Some students, e.g., S6, S12 \& S27, expressed their concerns towards danmaku in that sometimes sending danmaku could be contagious, and lots of irrelevant and useless danmakus might \emph{“interrupt the class and distract people’s attention”} (S6). This issue could be prevented if users clicked the "disable danmaku" button. For example, S29 chose to disable danmaku during one of her classes when \emph{“the instructor said something very funny, and everyone started to send `haha' or amusing emojis”} to avoid distraction. Another concern is that danmaku cannot be saved, but sometimes \emph{“the danmaku content can be essential to learning and worthy of being saved”} (S18). Similar perspectives were also expressed by S6 and S8. 

Unlike students who were enthusiastic about danmaku, teachers rarely used it. As T2 commented, \emph{“danmaku is more of a communication tool between students than between us and students”}, and T1, T2, and T3 each reflected that they rarely checked danmaku or did not check it at all. Moreover, although T4 always paid attention to danmaku during class since there were some \emph{“valuable questions”}, he mentioned a drawback of danmaku that \emph{“sometimes students send long and complicated danmaku, but those danmakus just slipped away so quickly that I cannot catch them”}.

\textbf{\textit{Quiz.}}
Quizzes were occasionally assigned to students during LS classes~(see Figure~\ref{fig:quiz} as a demonstration). Most instructors used the quiz to test \emph{“whether students paid attention in class”} and to \emph{“get some feedback from students”} so that they could adjust their teaching schedules accordingly (T2). T4 also mentioned that quizzes could remind students to focus on the class if they got distracted. Correspondingly, from students’ point of view, e.g. S6 \& S16, solving quizzes was indeed \emph{“helpful in maintaining concentration in class”} (S6). Moreover, some student interviewees, including S18 \& S20, also agreed with the instructors that \emph{“quizzes were an efficient way to reflect on our [the students’] understanding of the materials so that the instructor can adjust the pace and materials based on quiz results”} (S20). 

A quiz consisted primarily of simple multiple-choice questions. Some instructors mentioned in the interview that they would have liked to try some short-answer questions, but this would have caused technical issues, including inaccurate \emph{“character recognition”} (T2). T4 also reflected that most quizzes could not \emph{“reflect students’ understanding comprehensively”} due to their limited length and form. 

Technical instability was reported by students while taking quizzes. Since in most cases there would be a grade for the quiz which might affect one's grade point average (GPA), technical issues could significantly affect students’ grades. According to S7 and S8, some of their classmates encountered technical issues with the quizzes: \emph{“there was one time when my friend could not submit the quiz due to a technical software issue and got a zero on it”} (S7). Furthermore, many students reported that taking classes remotely made them lose focus on class, which consequently affected their performances on quizzes: S5 mentioned that sometimes he \emph{“made many mistakes on simple questions when I [he] felt distracted and asleep”}. 

\textbf{\textit{Vote.}}
During live streaming sessions, the instructor would sometimes use votes (as shown in Figure~\ref{fig:quiz}) to better understand students. Unlike taking quizzes, votes would not generate scores or affect students’ grades. For instance, S25 mentioned that his teacher had initiated a vote to choose \emph{“submitting an electronic version of homework or taking pictures and then uploading them”} (S25), and S29 reported that in one of their art history classes, the instructor had initiated a vote on students’ preferences on works of art. Based on the vote, the instructor would focus on the most favored ones during the class, which \emph{“makes the materials of the class more fascinating and humanized”} (S29). Moreover, some interviewees, e.g. S6 and S8, reflected that votes \emph{“encouraged more participation in the class, inspired more discussions, and thus made us [students] feel more focused and energetic”}.

\subsection{Individual Learning Experience Under LS Learning}

Besides specific interaction features, students and instructors also illustrated how students' overall learning experiences were shaped by live streaming-based remote education under the COVID-19 pandemic. Here three major aspects of individual learning experience arise: study time, focus and engagement, and learning outcomes.

\textbf{\textit{{\color{blue}Similar or longer study time.}}}
Some students, including S5, S9, S10, S16, S21 \& S25, reported that they spent approximately the same amount of time on studies compared to the pre-pandemic traditional learning. As reported by S21, \emph{“we just followed the same schedule before the pandemic outbreak, so there were not many differences: same classes, same workload, same people”} (S21). However, many other students in our interview study reported that they were studying longer time under LS learning.
Many of these interviewees attributed this to decreased study efficiency (e.g., S2, S3, S6, S14, S20). As reported by student participants, their study efficiency was lowered due to the decrease in concentration (S2, S3, S6, S18, etc.). We would talk about the reasons for students' decreasing concentration during LS learning in more detail in Section 4.2.2.


Some interviewees also reflected that they had spent more time searching for extra-resources (S29), reviewing class recordings (S2, S6), and doing homework, since peer collaborations and discussions, which benefit a lot for course understanding and homework completion, were hard to organize online (S11 \& S29). This increase in study time per course was further corroborated by our survey: 37\% of students reported an increase in time on average for each course; in comparison, 22\% of students reported a decrease. 

Besides self-study time, it is also noted that non-study issues took up much time that should have been dedicated to studies during live streaming sessions {\color{blue}from the interview}. Several students, including S3, S18, S20, S23, and S27, reported that they had spent less time on actual studying in lectures because of technical difficulties. According to S23 and S27, there was lots of in-class time wasted on adjusting equipment and questions such as asking whether others could hear them. This cut into the amount of actual learning time during the lecture.


\textbf{\textit{{\color{blue}Decreased focus and engagement.}}}
Almost all interview participants reported low study efficiency at home. They found it difficult to focus on live streaming videos, and this reduced their studying productivity. As commented by S10, \emph{“I kept telling myself that I should be productive and focus on the instructor, but I always got distracted by some trivial stuff while staring at the screen and when I realized I was not paying attention to the class, I was already far behind ... so I just gave up on this class and did my own stuff”}. Many student interviewees, including S1, S3, S4, S7, S14, S18, and S25, expressed experiences similar to those of S10’s. Our survey also supported this conclusion: 43\% of students reported a decrease in concentration, but only 27\% of students reported an increase.

{\color{blue}Much of this phenomenon has been attributed to the change in the environment as indicated by our interviews}: from campus, a community-based public area consisting of various facilities such as libraries, dormitories, and classrooms, to home, an individual private area with limited resources. The absence of peers physically around and the lack of a community-based learning environment were mentioned as two leading factors in the decreased focus, engagement, and productivity. \emph{“At my own home, I was just not in the mood to study”}, said S29, \emph{“and I really missed libraries since looking at peers studying in libraries would motivate me to study”}. S7 also conveyed the same feelings, \emph{“in libraries there are a lot of students studying, but now there is only me, which makes me feel very lonely”}. Moreover, according to S25, \emph{“since there were no people around me, I did not feel any pressure to study”}, and thus she felt she could \emph{“do whatever I [she] want[s] and allow[s] myself [herself] to get distracted”}. The survey also reflected similar results that the mean score of whether remote learning created a good intellectual environment decreased from 3.14 in Fall 2019 to 3.07 in Spring 2020. A drop of 0.07 supported the finding that a less satisfactory learning environment during remote learning. However, this drop did not reach 1/3 of the standard deviation and we concluded that not a meaningful educational effect was spotted through our survey. 

Multitasking~\cite{czerwinski2004diary,cao2021large} was a leading consequence of the changes in the study environment. Without the instructor's supervision, many students {\color{blue}from the interview} reported that they spent most of their time playing on their phones. Conversely, in a traditional classroom, students admitted that they would not have used their phones as frequently since \emph{“obviously the instructor could see me [them]”} (S2). In addition, S9 described another common multitasking situation: \emph{“I used to listen to peers carefully when they would ask or answer questions; however, now I always browse websites while other students are talking”}, and situations such as eating (S11) or even sleeping (S15) while taking LS courses are also mentioned. 

Apart from multitasking, interruption was another consequence of the changes in the study environment. Many {\color{blue}interviewees} said that there was always trivial stuff happening at home which would interrupt the ongoing lecture. For example, \emph{“sometimes there were people knocking at the door so I need to open the door”} (S27) or \emph{“my mom suddenly asked me to help her move the table”} (S21). Consequently, those domestic interruptions strongly distracted students’ attention from the class. 

In addition, fatigue was also a common problem with learning from home. The majority of the {\color{blue}interviewees}, including S5, S6, and S8, reflected that after staring at the screen for a long time, they felt dizzy, exhausted, and heavy-eyed. As S5 described, \emph{“on Tuesday, I have classes from 8 am to 12:15 pm, then have lunch, and then take classes from 1:30 pm to 3:05 pm. During this process, I just feel so dizzy staring at the screen in my room. (I) could not concentrate. However, when I looked at the blackboard in a real classroom, I do not feel this tired ... I am supposed to do homework in the evening, but I just don’t feel like studying or doing any work. And days just pass like this. Very inefficient and wasted. I also don’t get the chance to talk to people (in real person), to go out for a walk, or to do sports, which should ameliorate fatigue.”} (S5)

\textbf{\textit{Learning outcomes.}}
In terms of learning outcomes, not only grades but also knowledge mastery through theory and practice and social skills are investigated so as to better understand students' overall learning performances. From the perspective of grading, most {\color{blue}interviewees} reported that their grades had been affected little by the switch to LS class (e.g., S2, S4, S5, S6, S7, S11 \& S12). As S21 explained: \emph{“The instructor teaches the same materials as before, and the only difference is that now we take virtual classes”}. The result of the survey was also parallel to S21's comment. According to the survey, students gave relatively similar scores on whether instructors had a clear and helpful course structure (AVE(Fall 2019)=3.21 vs. AVE(Spring 2020)=3.17) and whether the course developed critical thinking during remote learning (AVE(Fall 2019)=3.21 vs. AVE(Spring 2020)=3.20) on average. 

Several {\color{blue}interviewees}, e.g., S8, S14 \& S16, shared that their grades had been higher than before because they felt \emph{“the instructors were less harsh on grades and gave relatively higher GPA”} (S16). When we interviewed teachers about the grading policies, some instructors, including T2, T3 \& T5, did acknowledge that they had given open-book exams and were more generous in grading due to online learning. Some other students attributed higher grades to the more effective virtual teaching environment. According to S1 and S8, they got higher grades since they felt live streaming classes made them \emph{“listen to the instructor more clearly”} (S8) and \emph{“there would not be any noises or chatting by students during a live streaming class”} (S1). A few {\color{blue}interviewees} reported lower grades because of remote education due to the change in environment, as was explained in detail in Section 4.2.2. \emph{“I just did not have a feeling of studying”}, S3 commented, \emph{“since there was no peer pressure to motivate me to study”}. 


From the theoretical perspective of knowledge mastery, there seems to be little difference between traditional learning and LS learning. The survey suggested similar scores on how well students had mastered field-specific knowledge and skills: a drop from 3.31 in Fall 2019 to 3.27 in Spring 2020 was spotted, creating a small difference that did not reach 1/3 of the standard deviation and thus not showing a meaningful effect. Similar perspectives were also expressed in the interviews. The majority of interviewed students reported that they did not think they had mastered more or less material than they had in traditional learning (e.g., S1, S8, S18, S23 \& S27) since \emph{“the materials that were required to understand did not change”} 
(S27). 

Besides theories, practice was also essential to the mastery of course materials. Many students complained about their laboratory courses or courses that involved real-world practices. As the survey indicated, the average score of whether remote learning promotes research experiences dropped from 2.31 in Fall 2019 to 2.15 in Spring 2020. With a drop of 0.16 that exceeds 1/3 of the standard deviation, a meaningful education effect was observed. This is corroborated by the interview study as well. For example, S4, S5, S6, S16, and S27 all studied engineering-related majors. All of them mentioned they had struggled while conducting experiments remotely. S5 complained that since it was impossible to conduct experiments remotely, \emph{“the teacher sent us [students] a photograph of the apparatus and asked us [students] to draw pictures based on the photograph ... the photo was too vague to reflect the 3D structure in detail ... we could only use imagination to draw”}. S20 also pointed out that in one of her classes, students were supposed to provide legal aid to people, but this real-world practice of law was canceled and changed to online case reviews. \emph{“The whole class just became pointless”}, S20 commented. Some students reported good online practice. For example, S29 reflected: \emph{“Live streaming sessions actually brought new forms and inspired students. For example, in one of my modern art history classes, there was a presentation session. Normally in a traditional class, we would just make a PPT for the presentation. However, in a live streaming presentation session, many students were inspired to use more innovative forms, such as animation and mindmap.”} 

Other than academic aptitude, socializing was also important to college students to maintain mental health~\cite{moeller2019loneliness}. Many {\color{blue}interviewees}, including S21 \& S27, expressed that they were afraid of the gradual loss of in-person communication ability, if remote classes were instituted as a long-term policy. As reported by S21, \emph{“although we could chat through audio or some chatting software, that kind of daily casual interaction kind of disappears”} (S21). S29 agreed with S21 and added that he felt due to the loss of daily casual interaction with peers, \emph{“it was kind of ‘cheaper’ to communicate online, and I felt I was getting lazy and gradually losing the ability to make real in-person conversations with people”}.  
The results of the interviews were parallel with those of the survey. From the perspective of students, the mean score of social skills dropped from 3.19 for Fall 2019 semester to 3.03 for Spring 2020 semester, which makes a decrease of 0.16 that surpasses 1/3 of the standard deviation and thus suggests a meaningful educational effect. 



\subsection{Collaborative Learning Experience Under LS Learning}
In terms of collaborative learning experiences under LS learning, three kinds of experiences emerge: 1) the instructor-student interaction experience, 2) the student-student interaction experience, and 3) the sense of community.

\textbf{\textit{Instructor-student interaction: most experience closer relationships.}}
According to the survey, the average interaction score between teachers and students, from the students' perspectives, increased from 3.19 in Fall 2019 to 3.22 in Spring 2020, where the increase does not reach 1/3 of the standard deviation and thus does not demonstrate a meaningful education effect. However, in our interview study, many students did report that they had developed a closer relationship with their instructors. S7, S8, S16, S18, S20, S25, S27, and S29 each mentioned that live streaming-based classes, especially the text box and the danmaku, had made them feel \emph{“more relaxed, less awkward, and more motivated to respond to teachers”} (S29). \emph{“I really like the text box”}, S7 said, \emph{“since we can comment or ask questions whenever we want without interrupting the class”}. Similarly, S25 commented about the danmaku modality that \emph{“we don’t need to raise our hands and wait to be called anymore ... we can always send a danmaku whenever we want to say something”}. Many students, including S8, S27 \& S29, also reflected that the anonymity of danmaku encouraged more students to participate in class, and thus \emph{“shorten[ed] the distance between teachers and us [students]”} (S20). Furthermore, S29 not only reported more frequent conversations with the instructor in class, but also after class: \emph{“due to live streaming classes, I found some of my teachers were actually very friendly and interesting, and I had not felt this way before. Sometimes we would send funny danmaku during the class, and the way our instructor responded was as if he was our friend. After class, I would directly send messages to the instructor if I had questions.”} (S29)

A few {\color{blue}student interviewees} reported an equal or a more distant relationship between themselves and their teachers. S9 and S21 felt that their relationships with teachers had become neither closer nor more distant since \emph{“we [students] were having classes just as before and we [students] never felt close to teachers ever”} (S21). For some students, their perceptions of instructor-student relationships may have changed over time. For example, S10 emphasized that in the beginning of the live streaming remote education, he had not been used to having conversations with teachers through online chatting, so he \emph{“felt distant with the teachers in the beginning”} (S10); but soon he acclimated to the new learning mode and no longer felt the same way anymore. Moreover, S8 mentioned that although he felt \emph{“psychologically closer to the instructor”}, he also felt \emph{“physically distant from the instructor”} (S8). 

Contrary to the closer student-instructor relationship reported by most students, some teachers felt that interaction with students was less efficient and that they had a more distant relationship with students because they could not see students' faces during a live-streaming session. As we explained in Section 4.1.2, in most cases students would turn off their videos due to network instability. Many instructors mentioned in our interview that the inability to see students’ faces during LS classes would impede \emph{“some of the emotional attachment”} and \emph{“instantaneous feedback”} between teachers and students (T4). As T4 reflected: \emph{“As a teacher, I can see whether students understand the materials through their eyes and their stares. If many students look really confused, I can immediately receive this feedback from their faces and explain the material again. In addition, recognizing students’ faces is significant to me since it makes me feel close to students emotionally”.} (T4) However, this was relatively hard to maintain when courses are turned online.

Furthermore, according to most interviewed instructors, this loss of instantaneous instructor-student feedback decreased teaching efficiency. For example, T1 described: \emph{“I could not see students, and I did not know what they were doing. They might have been playing computer games or chatting with friends while I was teaching. They could have been doing whatever they wanted ... This is definitely a disadvantage of remote learning: significantly decreased teaching efficiency. A traditional class feels like a group discussion: both students and I contribute to the class; a remote class feels like me talking to myself. I don’t know how well students understood the materials or what the teaching pace should be.”} (T1)

Instructors tried various means to cope with this loss of real-time feedback: calling on students to answer questions (T1), using the chat box to answer students’ questions (T7), asking students to turn on their cameras occasionally (T1), and staying after classes or providing additional office hours sessions to talk to students (T6). In addition, some instructors did mention that some modalities of LS learning had indeed increased teaching efficiency. For example, T2 mentioned that giving quizzes during the class increased teaching efficiency since she could receive students’ grades immediately for a sense of their performance. These means did allow teachers to build closer relationships with students on average. According to the survey, the average score of interaction between teachers and students from the perspectives of teachers increased from 3.47 in Fall 2019 semester to 3.52 in Spring 2020 semester. However, the increase did not prove to hold a meaningful education effect since it did not reach 1/3 of the standard deviation.

\textbf{\textit{Student-student interaction: most experience more distant relationships.}}
In terms of interactions between students, our survey result suggested a decrease in student-student interaction. Students' average score on student-student interaction decreased from 2.88 for Fall 2019 to 2.74 for Spring 2020, where a drop of 0.14 exceeds 1/3 of the standard deviation and indicates a meaningful education effect. Although student-student experience involved various scenarios, participants in our interviews discussed this subject mainly through two lenses: studying and leisure interaction. In terms of students’ interactions during their study time, what was frequently mentioned is student-student interactions during peer collaborations. For the latter part, special attention was paid to activities unrelated to academics and social interaction between students during their leisure time. 

Many students reported less frequent interaction with other students and decreased work efficiency during online group work according to the survey: the mean score of students to report working effectively with each other decreased from 3.18 in Fall 2019 to 2.95 in Spring 2020, creating a meaningful education effect where the 0.23 drop exceeded 1/3 of the standard deviation. This is in line with what was reported in our interview study. S7 and S14 reflected that rather than actually collaborating, they simply split the work by assigning work to each group member, and during the process of group collaboration, there was no other communication. 
What's more, {\color{blue}our interview indicated} that some students may lack the motivation to attend online meetings and cooperate (S10 \& S14). For example, according to S10, there were always \emph{“some members who never showed up to the meeting and did not reply to any messages in the group chat”}. Although sometimes this is also the case for the offline settings, with a layer of screen mediated, cases like this are more likely to happen (S10 \& S14). Moreover, in a traditional meeting, students often booked a room so that all members could work together, where \emph{“each group member would know other members’ progress and give feedback immediately if something is going wrong”} (S10) and \emph{“a sense of collaboration was sensed”} (S15). However, when collaborations were turned online and peers were segregated physically, \emph{“not a sense that we are working together”} (S15) was felt. Students felt rather isolated and collaborations sometimes turned to the mere \emph{“assigning tasks to each individual and checking the progress of each other one by one”} (S10). 
In addition, since setting up a remote meeting for every group member was hard due to students’ different schedules, the chat box was frequently used to discuss ideas. In this way, group members would communicate asynchronously. S10 and S20 both disliked the deficiency of this asynchronicity, reporting that sometimes it was \emph{“hard to use text to convey the exact information”} (S20) and kind of \emph{“a waste of time while waiting for others’ responses”} (S20). 

On the other hand, some {\color{blue}student interviewees} reported an increase in student-student interaction, as well as increased study efficiency. As discussed in Sections 4.1.3 and 4.1.4, the chat box and danmaku stimulated a significant increase in communication between students, and some {\color{blue}student interviewees} reported \emph{“a deeper understanding of the materials”} {\color{blue} (S12)} and higher study efficiency since \emph{“teachers did not have time to answer every question immediately”} (S12) while other students can provide the expected answers in real-time. Furthermore, contrary to students who complained about the asynchronous communication of the chat box, S7 argued that this asynchronicity indeed \emph{“motivate[d] me to think more and think deeper before typing”}, thus \emph{“creating more accurate, useful, and efficient conversations”} (S7).   

In terms of students’ interactions during leisure time, as revealed by our interview, the majority of the students felt more distant in their relationships with peers due to the absence of in-person interactions. As S3 reflected on his feelings about taking a new class with no acquaintances: \emph{“I just don’t know how to start a conversation online. In a normal class, when I walk into the classroom and take a look at everyone, I will develop some senses of the peers: who they are, what are their majors, etc. But now, everyone is just a ‘name’ appearing on the screen. Before the remote learning, it was natural to get to know people and initiate some conversation: I would say ‘hi’ to people who were sitting near me or have natural conversations about the class. Later we would add each other on social media and maybe study together. This process went smoothly. However, remote classes make this process very weird and awkward.”} (S3)
Many participants agreed with S3, e.g., S2, S8, S9, S10, etc., that they had made fewer friends during live streaming classes than in the traditional face-to-face settings. Furthermore, as discussed in Section 4.2.3, some participants reported a decline of in-person social skills, which could be a serious problem in the long run. 

\textbf{\textit{Weaker sense of community.}}
{\color{blue}Sense of community~\cite{mcmillan1986sense,mcmillan1996sense} also stands out in terms of collaborative learning experiences, which pointed out that interaction is a critical factor of a community structure.} Given that most students reported weaker relationships with their peers, as stated in Section 4.3.2, it is no surprise that the majority of students reported a lower perceived class-based sense of community. As mentioned before, the lack of visual impressions and face-to-face interactions both contributed to weaker social ties between students. In our interview, S7, S9, and S10 all made similar comments: students \emph{“did not know what other peers or the instructor looked like”} (S7) and nor did they \emph{“talk a lot after class”} (S9). As S10 concluded, \emph{“although we had classes together, classmates seemed to be very far away, and I did not feel like I knew them, not to mention being close to them”}. A few {\color{blue}student interviewees} felt the opposite way. For instance, S16 reported a stronger class-based sense of community. According to S16, in a traditional class, \emph{“people just come to listen to the lecture and go”}, but in a live  streaming class, \emph{“everyone’s name is displayed on the screen”}. {\color{blue}In line with what has been manifested by Sun, Wang, and Carroll~\cite{sun2019distance}, this visibility of recognizable names kind of \emph{“reinforce(s) a feeling of connection between us”}.} 

To better understand the overall experiences of students, we turn to the result of our large-scale survey study. Comparing the experience of participation in class activities in Spring 2020 semester (AVE(Spring 2020)=3.60) with Fall 2019 semester (AVE(Fall 2019)=3.76), a decrease by 0.14 was spotted, which was around 1/3 of the standard deviation and thus could be considered as a meaningful educational effect. Therefore, we concluded that a weaker sense of community was experienced on average on the class level.

At the level of school-based sense of community, weaker school-based feelings of connection were reported by most student interviewees. The absence of an environment -- campus -- was the most primary reason. The interviews showed that physical presence on campus was essential to students’ feelings of connection to the university. For example, according to S6, \emph{“when I was on campus, every day I would go to the dining hall, the teaching building, the dorm, and the library. However, now the only space I can move between is from one room to another room at home”} (S6). Since the physical presence of campus strengthens students’ shared identity as part of a school-based community, its absence weakens this feeling of connection (S6). S12 and S23 both expressed similar feelings. 

Another important factor that contributed to the weaker school-based feelings of connection was the lack of campus activities. There was some evidence that connections with school-based communities “may arise as a feeling of organization commitment, which relates to people’s affinity to a group as a whole”~\cite{sun2019distance}. According to the interview, most extracurricular activities were canceled due to the closed campus. Even if some activities could happen online, according to S27, \emph{“many club members felt less motivated to organize online activities since they were normally not that attractive.”} Without those activities, many participants, such as S6, S7, S21 \& S27, reported a weaker or even zero sense of connection to the school. The survey also reflected that the lack of campus activities had a significant educational effect, since the score on participation in public activities decreased from 2.92 in Fall 2019 to 2.54 in Spring 2020, resulting in a decrease of 0.38 which was much higher than 1/3 of the standard deviation.

A few students {\color{blue}in our interview} reported no difference in school-based feelings of connection between taking traditional classes and having remote education. This happened especially when students had acknowledged that members allowed to participate in the LS classes had been limited to students from the same university. As S25 suggested, \emph{“although we did not know what they [other students] looked like during live streaming classes, we did know that they were from [university name]. Thus, I don’t think that live streaming classes lowered my sense of connection to [university name]”}. {\color{blue} This corresponds with Sun, Rosson, and Carroll's work~\cite{sun2018community}, where the common identity, in this case being a student at a certain university, is a core driver of connections among students.}

\section{Discussions}
\subsection{Instructors versus Students}
{\color{blue} As indicated by Stakeholder Theory~\cite{freeman1983stockholders,mitchell1997toward}, maintaining the balance between stakeholders' interests is essential for cooperation. When the scenario is specified to education, teachers and students act as the two main stakeholders, where past research has demonstrated that both consistencies and disparities can be identified between the two stakeholders~\cite{zheng2015understanding,zheng2016ask}. Our study echos these works, illustrating how experiences and perceptions of instructors and students can resemble and differ from the other.}

In terms of similarities, it can be derived from the interview study that teachers could understand students' dilemmas during online learning and would try their best to improve the existing issues of remote learning to further create a more ideal learning experience for students. Attention and performance were two leading examples. To prevent students' attention from being distracted, instructors tried to maintain students’ attention by interacting more with students through various means: calling out students to answer questions, giving online quizzes or voting, etc. To provide better LS learning experiences, instructors learned to use the modalities introduced in Section 4.1 especially for senior professors who were not familiar with the emerging platforms. From Section 4.2.3, we also noticed that the majority of instructors endeavored to accommodate students' needs during remote learning. For example, many teachers uploaded more supplementary resources to students and made virtual experiments possible or mailed the materials required for experiments to students to ensure that students get a good understanding as the traditional learning scenario.

However, discrepancies between instructors and students are shown, too. From the survey, we observed an obvious mismatch regarding the experience and quality of remote education reported by teachers and students. Specifically, although teachers had some sense of the difficulties that students might encounter learning at home, they were too optimistic about the extent of influence brought by the change of the environment. For example, students in the survey study reported a decrease from 3.22 in Fall 2019 to 3.00 in Spring 2020 on average regarding the entire online education experience, where the drop of 0.22 surpasses 1/3 of the standard deviation and thus creates a meaningful educational effect. However, teachers only reported a little drop from 3.41 to 3.40 from Fall 2019 to Spring 2020, the effect of which is far from being regarded as meaningful. The leading examples of teachers being overly optimistic about the impacts brought by remote learning on students included the importance of an effective learning environment, the experience of conducting researches or practices remotely, the interactions between students, etc. 

Another major gap between instructors and students related to instructor-student interaction in LS learning. According to the interview, almost all instructors expressed their concerns about not seeing students' faces during live streaming sessions. As Section 4.3.1 concluded, many instructors reported less effective interactions and more distant relationships with students because of the loss of non-verbal expressions. However, from the perspective of students, non-verbal expressions were not mentioned at all. As reported by students, not showing faces to instructors did not influence their perceptions of relationships with instructors, and turning on their front cameras would only help them concentrate on class better. 

\subsection{Blur of Study and Life}
{\color{blue}The experiences of learning from home through LS learning has caused the blur of the boundary between studies and lives, which echos prior works demonstrating the impact of work from home on work-life boundaries \cite{jones2013work}.} Specifically, as we show in Section 4.2, most students' actions of learning are taken at home, sometimes even on the bed, and their only connection towards the school settings is the screen through which LS learning takes place. Therefore, they find it relatively hard to tell whether they are in a state of ease at home or they are in a tight mode at studies. In most cases, a blur and blend of the two is reported. What's more, under such a situation, it is also rather easy to switch between the status of studies and daily routines. In the absence of instructors' timely monitoring, some students have a higher tendency of multitasking while taking courses -- having breakfast, playing with their cell phones, etc., as demonstrated in Section 4.2.2, which coincides with remote meeting multitasking behavior \cite{cao2021large}. {\color{blue}In line with similar scenarios for MOOCs~\cite{xiao2017undertanding},} this has led the learning process to be more casual and lowered their learning efficiency. Their learning processes are also more likely to be interrupted by daily necessities at home, where their parents may have them do housework, greet guests to their houses, etc. This may turn learning to be somehow intermittent and lead students to miss some parts of the courses, which is detrimental to the efficiency and effectiveness of studies. 

\subsection{Decoupling Learning from Home and Learning Enabled by Live Streaming}
Taken at home and enabled by live streaming, the LS learning experience under the COVID-19 pandemic share the features of learning from home plus learning enabled by live streaming. Here we set out to decouple how these two features shape LS learning experience during the COVID-19 pandemic, respectively. 

Learning from home saved time for commuting. However, {\color{blue}as shown in Section 4.2.1,} the total time for studies may not be reduced and sometimes may even increase. This is in part because learning from home also determines the context for learning, where the context of home calls for self-regulation and learning autonomy~\cite{weinstein1987fostering}. For rather self-disciplined students, their learning outcomes are intact or even improved because they can better manage the paces of their studies. However, for students who are less autonomous, this change of context can be detrimental to learning efficiency. Firstly, when the circumstance of classrooms at school is replaced by the comfortable home settings, as reported {\color{blue}in Section 4.2.1}, students become more relaxed and their study status is turned more casual. This can in turn lead students to be more likely to be distracted and reduce the efficiency of learning. Secondly, the context of home reduces the distance between one's desk for learning and bed and kitchen, which provides prerequisites for students' multitasking such as sleeping or eating while course taking. When students are in lack of self-discipline, the effectiveness of education would be impaired. Thirdly, the aforementioned interruptions {\color{blue}mentioned in Section 4.2.2} are also somewhat home-specific, where forces to greet guests and disturbances by other family members are seldom the case for other scenarios.

Live streaming enables features that facilitate efficient and engaging learning. Firstly, the support of multiple modalities enriches the channels for interactions. {\color{blue}Past researches have shown how the incorporation of different interaction formats would benefit learning (e.g., \cite{hamilton2018collaborative,poon2019engaging,chen2019integrating}). Extending these works, we discover that the various modalities enabled by LS can encourage students to more willingly participate in the course and promote better instructor-student communications and interactions in formal education settings.} As reported by our interviewees {\color{blue}in Section 4.3}, this can in turn bring the relationships and perceived distances between instructors and students closer regardless of the physical segregation, which contributes to better study experiences. Secondly, LS learning is praised for its synchronous nature by LS learning practitioners. The incorporation of live streaming not only makes it possible for timely Q\&A and feedbacks {\color{blue}as mentioned in Section 4.1}, but also creates an atmosphere that the class is specially for the students (especially when compared with MOOCs), which, as articulated by LS learning students, increases their enthusiasm for participation and engagement. {\color{blue}This provides further evidence for the engaging~\cite{hamilton2014streaming,lu2018you}, immediate~\cite{haimson2017makes,lessel2017expanding}, and interacting~\cite{haimson2017makes,lu2019vicariously,lu2019feel} nature of live streaming emphasized by previous literature.}

\section{Implementations, Implications and Future Takeaways}
In this section, we present the practical implications of our research, including guidelines for lectures under LS learning, design implications to better support LS learning, and future takeaways from LS learning in hybrid education experiences.

\subsection{LS Learning Implementations for Different Courses and Disciplines}

As revealed in Section 4, the adaptability of face-to-face education to LS learning may vary across courses and disciplines. Here we aim to identify the feasible implementations of live streaming based remote education from the angle of course characteristics. {\color{blue}In terms of implementations for different courses, based upon instructionism and constructionism~\cite{sawyer2005cambridge}, we here discuss the concrete implementations where 1) the transmission-and-acquisition style lecture of instructionism hold: large-scale and small-scale lecture courses; 2) the interactive constructionism pedagogy is essential: interactive courses; and 3) hands-on practices are underlined: practice courses and labs.}

\textbf{\textit{Large-scale lecture courses.}} Numerous courses, especially general foundation courses and engineering courses, take the form of large-scale lectures, where the top priority is the transfer and acquisition of knowledge {\color{blue}as emphasized by instructionism~\cite{sawyer2005cambridge}}. In these courses, not much difference between face-to-face learning and LS learning is perceived in terms of outcomes (T4). Sometimes when the scale of the lecture classes is sufficiently large, the learning experiences may be even better without disturbance and with better sights of course contents (S1, S8 \& S13). In most cases of this scenario, the mere sharing of the instructors/lecturers' audio, video, and screen would be sufficient to enable quality studies. Students' basic interactions such as Q\&A and greetings with the instructors can be satisfactorily accomplished by text box messages and danmakus, and learning status and effectiveness can be timely checked through quizzes. However, if group presentations are integrated, the sharing of students' audio, video, and screen would be necessary. 

\textbf{\textit{Small-scale lecture courses.}} Similar to large-scale lecture courses, the combination of textual messages, instructors' audio, video, and screen sharing would support the basic delivery of course contents of small-scale lecture classes, where text box messages and danmakus support in-time interactions. One way for small-scale classes to motivate engagement is to enable direct verbal Q\&A where instructors are convenient to ask anyone and to directly check how someone learns. However, if this function is to be supported, LS learning platforms should allow two-way mutual audio interactions, where students should be allowed to turn on their microphones to voice their opinions and answers.

\textbf{\textit{Interactive courses.}} Some courses call for the recurrent instructor-student and student-student interactions {\color{blue}highlighted by constructionism~\cite{sawyer2005cambridge}}, for example, oral interpretation, case study, and group discussion classes. If the instructor needs to frequently check students' speakings, mutual verbal interactions between instructors and students are indispensable. If frequent group discussions are a must, it is anticipated that LS learning platforms should pay special attention to improving their backing on grouping. Therefore, diversified forms of interactions including exchanges of audio and video information between multiple people are in demand. What's more, to guarantee better experiences and effectiveness of peer discussions, one possible solution would be to support the sharing of everyone's images and turn on everyone's camera (at least in the group) so as to avoid awkwardness and unfamiliarity. 

\textbf{\textit{Practice courses.}} Practice courses such as dancing, painting, and physical education classes~\cite{cloudpe} call for frequent visual monitoring and timely guidance and corrections. Therefore, support for both instructors' and students' videos is crucial, and sometimes clarity and smoothness is highly required in terms of the video quality (S2 \& S29). Delay may be detrimental because if the timeliness for information to be passed through the instructor-student-instructor cycle is not guaranteed, the instructions given by the instructors may run behind what the students are actually operating, which would reduce the virtue of the instructions. Therefore, timely feedbacks and strong synchronization are vital.

\textbf{\textit{Labs.}} Labs are relatively the hardest to implement through LS learning. However, some solutions may also be provided for this form of courses to be conducted remotely through a live streaming format~\cite{cloudlab}. When the materials are convenient for mailing, sending the objects directly to students and letting students remotely follow the instructors at home would be preferable (S19). When the equipment is large and expensive but can be directly connected to computers, changing the experiments to a computer-mediated version would be an option (T5). Through remote control on the computers that instruments are connected to, the experiments can be accomplished remotely, which also allows the successful progression of labs.

{\color{blue}Implementations of LS learning also vary across different disciplines. For example, for science, technology, engineering, and mathematics (STEM) subjects, the vast majority of professional courses take the form of lectures (S15), where instuctionism-style implementations would be sufficient to meet the primary needs of knowledge transfer and acquisition. Meanwhile, experiments and practices are required for concrete mastery of the knowledge (T5), where the aforementioned implementations for hands-on practices should be taken into consideration. For art and humanity subjects, reflections, discussions, and group seminars are indispensable for most courses and the interactive constructionism-style implementations would be strongly preferred (S9, S18, T1 \& T7). }

\subsection{Design Implications}
Our work provides novel design implications for the HCI and CSCW community. We demonstrate users' usages and perceptions on different interaction features in LS learning, which can indicate the design of future platforms to support live streaming based education of the kind. Specifically, to improve the user experiences in LS learning and to improve the applicability of LS learning platforms, we delineate the typical features of promising LS learning platforms. 

\textbf{\textit{Diverse interaction formats.}} It is recommended that LS learning platforms should support diverse forms of interactions, including but not limited to audio, video, text box messages, danmakus, quizzes, and votes. {\color{blue}In line with prior works~\cite{hamilton2018collaborative,chen2019integrating}}, we discover that it would be plausible to enable two-way audio, video, and screen sharing to address the demands of different courses. {\color{blue}However, our study showcases that beyond these formats, the inclusion of emerging modalities such as danmakus and education-specific modalities such as quizzes and votes would also be beneficial.} Text box messages and danmakus provide two ways of textual engagements where the former is more formal while the latter is perceived to be more relaxing and casual. Quizzes grant prompt feedbacks of students' learning, while votes simplify thought collection and encourage engagement. 

\textbf{\textit{Balance of anonymity and real-name system.}} While the real-name system of text box messages is appreciated, the anonymity of danmaku is also welcomed. {\color{blue}Therefore, for interface designers, it is worth carefully considering whether real-name systems are integrated and to what extent and for which functions real-name systems and anonymity may be used to balance engagement, content quality, and social relationships~\cite{ma2016anonymity, whiting2020parallel, cao2020my}.} For example, a combination of real-name systems and anonymity may be a feasible choice. With different systems integrated into different modalities, one can find a channel suits better for him/her and for his/her words.

\textbf{\textit{Activation and deactivation.}} To avoid disturbance, we advocate that instructors should have the right to decide if features such as danmuku and students' sharing of their voices and visuals would be activated through the courses, and students should be allowed to show or hide the messages and danmakus. From the instructor's perspective, it is up to him/her to decide how the course progresses and if an interaction format is allowed; from the student side, it would be better if he/she is allowed to adopt a learning circumstance that suits him/her best.

\textbf{\textit{Focus mode.}} {\color{blue}Existing work showcases that enabling viewpoint sharing, which act as a sort of visual focus mode, would benefit participation~\cite{hamilton2018collaborative}. Extending the past literature, we discover that not only visual but also audio focus mode would be appreciated by LS learners.} Specifically, as mentioned by interviewees in Section 4.1.1, if someone else keeps his/her microphone on during lectures, the fluency of the instructors' speeches received by students would deteriorate, where the students' voices would be very bothering. Supporting a focus mode would solve the problem to a certain degree. It would be intriguing if users can decide their priorities for accepting whose audio and video information. If one is allowed to prioritize the acceptance of the instructors' voices and visuals, other students' unintended vocal disturbance would not hinder his/her studies.

\textbf{\textit{Hierarchical role system.}} LS learning instructors and students also appreciate the support of hierarchical role systems, which echos the case of previous work such as Hamilton et al.'s~\cite{hamilton2018collaborative}. Instructors should maintain control of the whole course, while mediators (maybe teaching assistants and group leaders) can be assigned some of the instructors' rights so as to better regulate the course. What's more, with some of the rights only attributed to certain roles such as group leaders, less bandwidth would be needed for allocation and the robustness for disturbance is enhanced.

\textbf{\textit{Peer work and collaboration.}} Support for peer work and group collaboration is warmly welcomed. {\color{blue}This corroborates the necessities of the explorations on peer learning and remote collaboration in the education settings (e.g.,~\cite{kotturi2015structure,cheng2019teaching}).} We find that instructors and students speak highly of the `group discussion' function, where both random grouping and assigned grouping have their own merits. What's more, instructors such as T1 call for the simultaneous monitor of different groups so as to better control the progression of the whole class.

\textbf{\textit{Recordings and playbacks.}} Course recordings and playbacks have been highly praised by most interviewed students. This is perceived to contribute to a better understanding of the course contents and benefit the review of lessons. For example, students are likely to be distracted due to factors like interruptions. If recordings and playbacks are allowed, LS learning can absorb the convenience of pausing, rewinding, and reviewing~\cite{zheng2015understanding,kovacs2016effects}, where students can replay the parts they miss or fail to understand and thus achieve better comprehension. However, concerns on course recordings are raised by instructors. Not only are they afraid of copyright infringement and vicious dissemination of certain contents (T1) {\color{blue} as indicated by previous studies on video privacy~\cite{senior2005enabling,crow2017community}}, but they also express concern that too much reliance on playbacks would lead to laziness: not keeping up with the class and only watching all the playbacks before exams, which would cause a significant drop in the effectiveness of learning (T7).

\subsection{Future Takeaways}
Discussions are also made on what can be derived from the experiences of live streaming based remote education during the specific period of the COVID-19 pandemic. Here we demonstrate the future takeaways that can be extracted for post-pandemic education. 

{\color{blue}Firstly, experiences from LS learning shed light on the promising future of hybrid learning, where the advantages of online and offline learning may be integrated in a similar way as `hybrid work'~\cite{halford2005hybrid}.} Specifically, there is the possibility that certain interaction features of LS learning may be kept for utilization in post-pandemic periods. For example, the allowance of chatbox and danmaku may enhance instructor-student interactions and benefit the sharing of different perspectives which may contribute to a more active class atmosphere. The usage of quizzes and votes would enable instructors to have first-hand knowledge of students' current knowledge mastery, which helps instructors to accordingly adjust their teaching in time. {\color{blue} With related areas demonstrating the effectiveness of real-time text chat~\cite{jones2008empirical,hamilton2014streaming}, live stream~\cite{haimson2017makes} and in-course quizzes~\cite{kovacs2016effects, poon2019engaging}, it is reasonable to anticipate improved learning experience of traditional classes through the incorporation of diverse modalities. Hamilton et al.'s work~\cite{hamilton2018collaborative} would be a representative attempt, where there is still room for further improvements for the incorporation of further features such as quizzes.} 

Secondly, LS learning demonstrates the possibilities of effective distant education. Although platforms such as MOOCs have been available for distance learning for years, the outcomes of the learning on those platforms can be far from satisfactory (S15), where \emph{"courses are seldom treated seriously"} (S21). However, {\color{blue}with engaging~\cite{hamilton2014streaming,lu2018you}, immediate~\cite{haimson2017makes,lessel2017expanding}, and interacting~\cite{haimson2017makes,lu2019vicariously,lu2019feel} live streaming,} LS learning provides a means with real-time interactions with the instructor, where students feel that \emph{"a live person is teaching for you"} (S14) rather than \emph{"I feel I am not taking a course at all"} (S11). Therefore, for future design targeting at improving the effectiveness of distance learning, {\color{blue}we advocate following the line of works utilizing live streaming~\cite{hamilton2014streaming, lessel2017expanding,lu2019feel,lu2019vicariously,lu2018streamwiki,yang2020snapstream,chen2019integrating} and considering LS learning.}

Thirdly, some defects still lie within LS learning which future HCI and CSCW researchers may take into consideration. For example, students feel it easy to be distracted because of a lack of formal study context. One possible direction for addressing the problem may be to set up a `real' virtual classroom space for interactions, possibly through technologies such as virtual reality (VR). {\color{blue}Existing works have already taken the first steps to enable virtual environments for education and the simulation of gestures and facial expressions for learning~\cite{greenwald2017technology}.} We call on future efforts to compensate for the defects reported by our participants while maintaining and even improving the advantages of LS learning through emerging technologies.

\section{Limitations, Generalizability, and Future Work}
Our research inevitably suffers from limitations. First of all, our data were collected in mainland China, so likely not all conclusions will apply to different cultures. For instance, aside from Zoom, other LS learning platforms, e.g., Tencent Meeting and Rain Classroom, are not widely used outside mainland China. Some modes of interaction, e.g., danmaku, are more East Asian specific. Therefore, one could expect different experiences on other platforms. Beyond that, culturally the instructor-student relationship in East Asia is more distant than in Western culture, so the implications of relationship may not naturally extend. Nevertheless, our study is based on large-scale interview and survey studies where we carefully sample participants from diverse backgrounds, which we argue ensures the generalizability of findings in China. In the future, we plan to extend our research to different cultures and disciplines, and carefully tease out the cultural effect on user experiences.

\section{Conclusion}
In this paper, we study live streaming based education experiences during the COVID-19 pandemic through mixed methods. With a focus on Chinese higher education, we carried out semi-structured interviews on 30 students and 7 instructors from diverse disciplines, meanwhile launched a large-scale survey covering 6291 students and 1160 instructors, and analyzed user experiences on LS education. Our findings suggest LS learning do help student and teachers achieve their education goal to a great extent under remote setting, yet there are several key challenges emerging under the current paradigm, including students' difficulties in paying continuous attention, decreased learning efficacy, and lack of engagement and collaboration. We further demonstrate how various interaction formats enable several novel learning experiences under LS learning, which contribute to variations in instructor-student and student-student relationships in both positive and negative ways. Based on our findings, we propose important design guidelines and insights to better support current remote learning experiences during the pandemic, and systematically discuss design implications to construct future collaborative education supporting systems and experiences post-pandemic.

\begin{acks}
This work was supported in part by The National Key Research and Development Program of China under grant 2020AAA0106000, the National Natural Science Foundation of China under U1936217,  61971267, 61972223, 61941117, 61861136003, Beijing Natural Science Foundation under L182038, Beijing National Research Center for Information Science and Technology under 20031887521, and research fund of Tsinghua University - Tencent Joint Laboratory for Internet Innovation Technology. Yu Zhang was supported by "Dual High" Project of Tsinghua Humanity Development titled "Tsinghua University Online Education Policy Evaluation during Epidemic Control".
\end{acks}

\bibliographystyle{ACM-Reference-Format}
\bibliography{sample-base}


\begin{thebibliography}{74}


\ifx \showCODEN    \undefined \def \showCODEN     #1{\unskip}     \fi
\ifx \showDOI      \undefined \def \showDOI       #1{#1}\fi
\ifx \showISBNx    \undefined \def \showISBNx     #1{\unskip}     \fi
\ifx \showISBNxiii \undefined \def \showISBNxiii  #1{\unskip}     \fi
\ifx \showISSN     \undefined \def \showISSN      #1{\unskip}     \fi
\ifx \showLCCN     \undefined \def \showLCCN      #1{\unskip}     \fi
\ifx \shownote     \undefined \def \shownote      #1{#1}          \fi
\ifx \showarticletitle \undefined \def \showarticletitle #1{#1}   \fi
\ifx \showURL      \undefined \def \showURL       {\relax}        \fi
\providecommand\bibfield[2]{#2}
\providecommand\bibinfo[2]{#2}
\providecommand\natexlab[1]{#1}
\providecommand\showeprint[2][]{arXiv:#2}

\bibitem[\protect\citeauthoryear{An, Bakker, Ordanovski, Taconis, Paffen, and
  Eggen}{An et~al\mbox{.}}{2019}]%
        {an2019unobtrusively}
\bibfield{author}{\bibinfo{person}{Pengcheng An}, \bibinfo{person}{Saskia
  Bakker}, \bibinfo{person}{Sara Ordanovski}, \bibinfo{person}{Ruurd Taconis},
  \bibinfo{person}{Chris~LE Paffen}, {and} \bibinfo{person}{Berry Eggen}.}
  \bibinfo{year}{2019}\natexlab{}.
\newblock \showarticletitle{Unobtrusively enhancing reflection-in-action of
  teachers through spatially distributed ambient information}. In
  \bibinfo{booktitle}{\emph{Proceedings of the 2019 CHI Conference on Human
  Factors in Computing Systems}}. \bibinfo{pages}{1--14}.
\newblock


\bibitem[\protect\citeauthoryear{Cao, Chen, Cheng, Zhao, Wang, and Li}{Cao
  et~al\mbox{.}}{2020a}]%
        {cao2020you}
\bibfield{author}{\bibinfo{person}{Hancheng Cao}, \bibinfo{person}{Zhilong
  Chen}, \bibinfo{person}{Mengjie Cheng}, \bibinfo{person}{Shuling Zhao},
  \bibinfo{person}{Tao Wang}, {and} \bibinfo{person}{Yong Li}.}
  \bibinfo{year}{2020}\natexlab{a}.
\newblock \showarticletitle{You Recommend, I Buy: How and Why People Engage in
  Instant Messaging Based Social Commerce}.
\newblock \bibinfo{journal}{\emph{arXiv preprint arXiv:2011.00191}}
  (\bibinfo{year}{2020}).
\newblock


\bibitem[\protect\citeauthoryear{Cao, Chen, Xu, Wang, Xu, Zhang, and Li}{Cao
  et~al\mbox{.}}{2020b}]%
        {cao2020your}
\bibfield{author}{\bibinfo{person}{Hancheng Cao}, \bibinfo{person}{Zhilong
  Chen}, \bibinfo{person}{Fengli Xu}, \bibinfo{person}{Tao Wang},
  \bibinfo{person}{Yujian Xu}, \bibinfo{person}{Lianglun Zhang}, {and}
  \bibinfo{person}{Yong Li}.} \bibinfo{year}{2020}\natexlab{b}.
\newblock \showarticletitle{When Your Friends Become Sellers: An Empirical
  Study of Social Commerce Site Beidian}. In
  \bibinfo{booktitle}{\emph{Proceedings of the International AAAI Conference on
  Web and Social Media}}, Vol.~\bibinfo{volume}{14}. \bibinfo{pages}{83--94}.
\newblock


\bibitem[\protect\citeauthoryear{Cao, Lee, Iqbal, Czerwinski, Wong, Rintel,
  Hecht, Teevan, and Yang}{Cao et~al\mbox{.}}{2021}]%
        {cao2021large}
\bibfield{author}{\bibinfo{person}{Hancheng Cao}, \bibinfo{person}{Chia-Jung
  Lee}, \bibinfo{person}{Shamsi Iqbal}, \bibinfo{person}{Mary Czerwinski},
  \bibinfo{person}{Priscilla Wong}, \bibinfo{person}{Sean Rintel},
  \bibinfo{person}{Brent Hecht}, \bibinfo{person}{Jaime Teevan}, {and}
  \bibinfo{person}{Longqi Yang}.} \bibinfo{year}{2021}\natexlab{}.
\newblock \showarticletitle{Large Scale Analysis of Multitasking Behavior
  During Remote Meetings}. In \bibinfo{booktitle}{\emph{the 2021 CHI conference
  on human factors in computing systems.}}
\newblock
\urldef\tempurl%
\url{https://doi.org/10.1145/3411764.3445243}
\showDOI{\tempurl}


\bibitem[\protect\citeauthoryear{Cao, Yang, Chen, Lee, Stone, Diarrassouba,
  Whiting, and Bernstein}{Cao et~al\mbox{.}}{2020c}]%
        {cao2020my}
\bibfield{author}{\bibinfo{person}{Hancheng Cao}, \bibinfo{person}{Vivian
  Yang}, \bibinfo{person}{Victor Chen}, \bibinfo{person}{Yu~Jin Lee},
  \bibinfo{person}{Lydia Stone}, \bibinfo{person}{N'godjigui~Junior
  Diarrassouba}, \bibinfo{person}{Mark~E. Whiting}, {and}
  \bibinfo{person}{Michael~S. Bernstein}.} \bibinfo{year}{2020}\natexlab{c}.
\newblock \showarticletitle{My Team Will Go On: Differentiating High and Low
  Viability Teams through Team Interaction}.
\newblock \bibinfo{journal}{\emph{Proc. ACM Hum.-Comput. Interact.}}
  \bibinfo{volume}{4}, \bibinfo{number}{CSCW3}, Article
  \bibinfo{articleno}{230} (\bibinfo{year}{2020}),
  \bibinfo{numpages}{27}~pages.
\newblock
\urldef\tempurl%
\url{https://doi.org/10.1145/3432929}
\showDOI{\tempurl}


\bibitem[\protect\citeauthoryear{Chen, Freeman, and Balakrishnan}{Chen
  et~al\mbox{.}}{2019}]%
        {chen2019integrating}
\bibfield{author}{\bibinfo{person}{Di Chen}, \bibinfo{person}{Dustin Freeman},
  {and} \bibinfo{person}{Ravin Balakrishnan}.} \bibinfo{year}{2019}\natexlab{}.
\newblock \showarticletitle{Integrating Multimedia Tools to Enrich Interactions
  in Live Streaming for Language Learning}. In
  \bibinfo{booktitle}{\emph{Proceedings of the 2019 CHI Conference on Human
  Factors in Computing Systems}}. \bibinfo{pages}{1--14}.
\newblock


\bibitem[\protect\citeauthoryear{Chen, Cao, Xu, Cheng, Wang, and Li}{Chen
  et~al\mbox{.}}{2020}]%
        {chen2020understanding}
\bibfield{author}{\bibinfo{person}{Zhilong Chen}, \bibinfo{person}{Hancheng
  Cao}, \bibinfo{person}{Fengli Xu}, \bibinfo{person}{Mengjie Cheng},
  \bibinfo{person}{Tao Wang}, {and} \bibinfo{person}{Yong Li}.}
  \bibinfo{year}{2020}\natexlab{}.
\newblock \showarticletitle{Understanding the Role of Intermediaries in Online
  Social E-commerce: An Exploratory Study of Beidian}.
\newblock \bibinfo{journal}{\emph{Proceedings of the ACM on Human-Computer
  Interaction}} \bibinfo{volume}{4}, \bibinfo{number}{CSCW2}
  (\bibinfo{year}{2020}), \bibinfo{pages}{1--24}.
\newblock


\bibitem[\protect\citeauthoryear{Cheng, Yu, Fu, Zhao, Hecht, Konstan, Terveen,
  Yarosh, and Zhu}{Cheng et~al\mbox{.}}{2019}]%
        {cheng2019teaching}
\bibfield{author}{\bibinfo{person}{Hao-Fei Cheng}, \bibinfo{person}{Bowen Yu},
  \bibinfo{person}{Siwei Fu}, \bibinfo{person}{Jian Zhao},
  \bibinfo{person}{Brent Hecht}, \bibinfo{person}{Joseph Konstan},
  \bibinfo{person}{Loren Terveen}, \bibinfo{person}{Svetlana Yarosh}, {and}
  \bibinfo{person}{Haiyi Zhu}.} \bibinfo{year}{2019}\natexlab{}.
\newblock \showarticletitle{Teaching UI design at global scales: A case study
  of the design of collaborative capstone projects for MOOCs}. In
  \bibinfo{booktitle}{\emph{Proceedings of the Sixth (2019) ACM Conference on
  Learning@ Scale}}. \bibinfo{pages}{1--11}.
\newblock


\bibitem[\protect\citeauthoryear{Coetzee, Fox, Hearst, and Hartmann}{Coetzee
  et~al\mbox{.}}{2014a}]%
        {coetzee2014chatrooms}
\bibfield{author}{\bibinfo{person}{Derrick Coetzee}, \bibinfo{person}{Armando
  Fox}, \bibinfo{person}{Marti~A Hearst}, {and} \bibinfo{person}{Bjoern
  Hartmann}.} \bibinfo{year}{2014}\natexlab{a}.
\newblock \showarticletitle{Chatrooms in MOOCs: all talk and no action}. In
  \bibinfo{booktitle}{\emph{Proceedings of the first ACM conference on
  Learning@ scale conference}}. \bibinfo{pages}{127--136}.
\newblock


\bibitem[\protect\citeauthoryear{Coetzee, Fox, Hearst, and Hartmann}{Coetzee
  et~al\mbox{.}}{2014b}]%
        {coetzee2014should}
\bibfield{author}{\bibinfo{person}{Derrick Coetzee}, \bibinfo{person}{Armando
  Fox}, \bibinfo{person}{Marti~A Hearst}, {and} \bibinfo{person}{Bj{\"o}rn
  Hartmann}.} \bibinfo{year}{2014}\natexlab{b}.
\newblock \showarticletitle{Should your MOOC forum use a reputation system?}.
  In \bibinfo{booktitle}{\emph{Proceedings of the 17th ACM conference on
  Computer supported cooperative work \& social computing}}.
  \bibinfo{pages}{1176--1187}.
\newblock


\bibitem[\protect\citeauthoryear{Cohen}{Cohen}{2013}]%
        {cohen2013statistical}
\bibfield{author}{\bibinfo{person}{Jacob Cohen}.}
  \bibinfo{year}{2013}\natexlab{}.
\newblock \bibinfo{booktitle}{\emph{Statistical power analysis for the
  behavioral sciences}}.
\newblock \bibinfo{publisher}{Academic press}.
\newblock


\bibitem[\protect\citeauthoryear{Corbin and Strauss}{Corbin and
  Strauss}{2014}]%
        {corbin2014basics}
\bibfield{author}{\bibinfo{person}{Juliet Corbin} {and} \bibinfo{person}{Anselm
  Strauss}.} \bibinfo{year}{2014}\natexlab{}.
\newblock \bibinfo{booktitle}{\emph{Basics of qualitative research: Techniques
  and procedures for developing grounded theory}}.
\newblock \bibinfo{publisher}{Sage publications}.
\newblock


\bibitem[\protect\citeauthoryear{Crow, Snyder, Crichlow, and Smykla}{Crow
  et~al\mbox{.}}{2017}]%
        {crow2017community}
\bibfield{author}{\bibinfo{person}{Matthew~S Crow}, \bibinfo{person}{Jamie~A
  Snyder}, \bibinfo{person}{Vaughn~J Crichlow}, {and}
  \bibinfo{person}{John~Ortiz Smykla}.} \bibinfo{year}{2017}\natexlab{}.
\newblock \showarticletitle{Community perceptions of police body-worn cameras:
  The impact of views on fairness, fear, performance, and privacy}.
\newblock \bibinfo{journal}{\emph{Criminal Justice and Behavior}}
  \bibinfo{volume}{44}, \bibinfo{number}{4} (\bibinfo{year}{2017}),
  \bibinfo{pages}{589--610}.
\newblock


\bibitem[\protect\citeauthoryear{Czerwinski, Horvitz, and Wilhite}{Czerwinski
  et~al\mbox{.}}{2004}]%
        {czerwinski2004diary}
\bibfield{author}{\bibinfo{person}{Mary Czerwinski}, \bibinfo{person}{Eric
  Horvitz}, {and} \bibinfo{person}{Susan Wilhite}.}
  \bibinfo{year}{2004}\natexlab{}.
\newblock \showarticletitle{A diary study of task switching and interruptions}.
  In \bibinfo{booktitle}{\emph{Proceedings of the SIGCHI conference on Human
  factors in computing systems}}. \bibinfo{pages}{175--182}.
\newblock


\bibitem[\protect\citeauthoryear{Dillahunt, Ng, Fiesta, and Wang}{Dillahunt
  et~al\mbox{.}}{2016}]%
        {dillahunt2016massive}
\bibfield{author}{\bibinfo{person}{Tawanna~R Dillahunt}, \bibinfo{person}{Sandy
  Ng}, \bibinfo{person}{Michelle Fiesta}, {and} \bibinfo{person}{Zengguang
  Wang}.} \bibinfo{year}{2016}\natexlab{}.
\newblock \showarticletitle{Do massive open online course platforms support
  employability?}. In \bibinfo{booktitle}{\emph{Proceedings of the 19th ACM
  conference on computer-supported cooperative work \& social computing}}.
  \bibinfo{pages}{233--244}.
\newblock


\bibitem[\protect\citeauthoryear{Dougherty}{Dougherty}{2011}]%
        {dougherty2011live}
\bibfield{author}{\bibinfo{person}{Audubon Dougherty}.}
  \bibinfo{year}{2011}\natexlab{}.
\newblock \showarticletitle{Live-streaming mobile video: production as civic
  engagement}. In \bibinfo{booktitle}{\emph{Proceedings of the 13th
  international conference on human computer interaction with mobile devices
  and services}}. \bibinfo{pages}{425--434}.
\newblock


\bibitem[\protect\citeauthoryear{Faas, Dombrowski, Young, and Miller}{Faas
  et~al\mbox{.}}{2018}]%
        {faas2018watch}
\bibfield{author}{\bibinfo{person}{Travis Faas}, \bibinfo{person}{Lynn
  Dombrowski}, \bibinfo{person}{Alyson Young}, {and} \bibinfo{person}{Andrew~D
  Miller}.} \bibinfo{year}{2018}\natexlab{}.
\newblock \showarticletitle{Watch me code: Programming mentorship communities
  on twitch. tv}.
\newblock \bibinfo{journal}{\emph{Proceedings of the ACM on Human-Computer
  Interaction}} \bibinfo{volume}{2}, \bibinfo{number}{CSCW}
  (\bibinfo{year}{2018}), \bibinfo{pages}{1--18}.
\newblock


\bibitem[\protect\citeauthoryear{Fang}{Fang}{2020}]%
        {cloudpe}
\bibfield{author}{\bibinfo{person}{Jia Fang}.} \bibinfo{year}{2020}\natexlab{}.
\newblock \bibinfo{title}{BBC Enters Tsinghua PE class}.
\newblock
  \bibinfo{howpublished}{\url{https://mp.weixin.qq.com/s/8JJP-5n6E6gYUXj_02Jx8w}}.
\newblock


\bibitem[\protect\citeauthoryear{Freeman and Reed}{Freeman and Reed}{1983}]%
        {freeman1983stockholders}
\bibfield{author}{\bibinfo{person}{R~Edward Freeman} {and}
  \bibinfo{person}{David~L Reed}.} \bibinfo{year}{1983}\natexlab{}.
\newblock \showarticletitle{Stockholders and stakeholders: A new perspective on
  corporate governance}.
\newblock \bibinfo{journal}{\emph{California management review}}
  \bibinfo{volume}{25}, \bibinfo{number}{3} (\bibinfo{year}{1983}),
  \bibinfo{pages}{88--106}.
\newblock


\bibitem[\protect\citeauthoryear{Fu}{Fu}{2020}]%
        {cloudlab}
\bibfield{author}{\bibinfo{person}{Peiyue Fu}.}
  \bibinfo{year}{2020}\natexlab{}.
\newblock \bibinfo{title}{Cloud experiments at home: No limits to learning at
  Tsinghua!}
\newblock
  \bibinfo{howpublished}{\url{https://mp.weixin.qq.com/s/rlAuhTpVLsEvx3FKiyK5MA}}.
\newblock


\bibitem[\protect\citeauthoryear{Greenwald, Kulik, Kunert, Beck, Frohlich,
  Cobb, Parsons, and Newbutt}{Greenwald et~al\mbox{.}}{2017}]%
        {greenwald2017technology}
\bibfield{author}{\bibinfo{person}{S Greenwald}, \bibinfo{person}{Alexander
  Kulik}, \bibinfo{person}{Andr{\'e} Kunert}, \bibinfo{person}{Stephan Beck},
  \bibinfo{person}{B Frohlich}, \bibinfo{person}{Sue Cobb},
  \bibinfo{person}{Sarah Parsons}, {and} \bibinfo{person}{Nigel Newbutt}.}
  \bibinfo{year}{2017}\natexlab{}.
\newblock \showarticletitle{Technology and applications for collaborative
  learning in virtual reality}.
\newblock  (\bibinfo{year}{2017}).
\newblock


\bibitem[\protect\citeauthoryear{Gumienny, Gericke, Wenzel, and
  Meinel}{Gumienny et~al\mbox{.}}{2013}]%
        {gumienny2013supporting}
\bibfield{author}{\bibinfo{person}{Raja Gumienny}, \bibinfo{person}{Lutz
  Gericke}, \bibinfo{person}{Matthias Wenzel}, {and} \bibinfo{person}{Christoph
  Meinel}.} \bibinfo{year}{2013}\natexlab{}.
\newblock \showarticletitle{Supporting creative collaboration in globally
  distributed companies}. In \bibinfo{booktitle}{\emph{Proceedings of the 2013
  conference on Computer supported cooperative work}}.
  \bibinfo{pages}{995--1007}.
\newblock


\bibitem[\protect\citeauthoryear{Haimson and Tang}{Haimson and Tang}{2017}]%
        {haimson2017makes}
\bibfield{author}{\bibinfo{person}{Oliver~L Haimson} {and}
  \bibinfo{person}{John~C Tang}.} \bibinfo{year}{2017}\natexlab{}.
\newblock \showarticletitle{What makes live events engaging on Facebook Live,
  Periscope, and Snapchat}. In \bibinfo{booktitle}{\emph{Proceedings of the
  2017 CHI conference on human factors in computing systems}}.
  \bibinfo{pages}{48--60}.
\newblock


\bibitem[\protect\citeauthoryear{Halford}{Halford}{2005}]%
        {halford2005hybrid}
\bibfield{author}{\bibinfo{person}{Susan Halford}.}
  \bibinfo{year}{2005}\natexlab{}.
\newblock \showarticletitle{Hybrid workspace: Re-spatialisations of work,
  organisation and management}.
\newblock \bibinfo{journal}{\emph{New Technology, Work and Employment}}
  \bibinfo{volume}{20}, \bibinfo{number}{1} (\bibinfo{year}{2005}),
  \bibinfo{pages}{19--33}.
\newblock


\bibitem[\protect\citeauthoryear{Hamilton, Garretson, and Kerne}{Hamilton
  et~al\mbox{.}}{2014}]%
        {hamilton2014streaming}
\bibfield{author}{\bibinfo{person}{William~A Hamilton}, \bibinfo{person}{Oliver
  Garretson}, {and} \bibinfo{person}{Andruid Kerne}.}
  \bibinfo{year}{2014}\natexlab{}.
\newblock \showarticletitle{Streaming on twitch: fostering participatory
  communities of play within live mixed media}. In
  \bibinfo{booktitle}{\emph{Proceedings of the SIGCHI conference on human
  factors in computing systems}}. \bibinfo{pages}{1315--1324}.
\newblock


\bibitem[\protect\citeauthoryear{Hamilton, Lupfer, Botello, Tesch, Stacy,
  Merrill, Williford, Bentley, and Kerne}{Hamilton et~al\mbox{.}}{2018}]%
        {hamilton2018collaborative}
\bibfield{author}{\bibinfo{person}{William~A Hamilton}, \bibinfo{person}{Nic
  Lupfer}, \bibinfo{person}{Nicolas Botello}, \bibinfo{person}{Tyler Tesch},
  \bibinfo{person}{Alex Stacy}, \bibinfo{person}{Jeremy Merrill},
  \bibinfo{person}{Blake Williford}, \bibinfo{person}{Frank~R Bentley}, {and}
  \bibinfo{person}{Andruid Kerne}.} \bibinfo{year}{2018}\natexlab{}.
\newblock \showarticletitle{Collaborative live media curation: Shared context
  for participation in online learning}. In
  \bibinfo{booktitle}{\emph{Proceedings of the 2018 CHI Conference on Human
  Factors in Computing Systems}}. \bibinfo{pages}{1--14}.
\newblock


\bibitem[\protect\citeauthoryear{Hauber, Regenbrecht, Billinghurst, and
  Cockburn}{Hauber et~al\mbox{.}}{2006}]%
        {hauber2006spatiality}
\bibfield{author}{\bibinfo{person}{J{\"o}rg Hauber}, \bibinfo{person}{Holger
  Regenbrecht}, \bibinfo{person}{Mark Billinghurst}, {and}
  \bibinfo{person}{Andy Cockburn}.} \bibinfo{year}{2006}\natexlab{}.
\newblock \showarticletitle{Spatiality in videoconferencing: trade-offs between
  efficiency and social presence}. In \bibinfo{booktitle}{\emph{Proceedings of
  the 2006 20th anniversary conference on Computer supported cooperative
  work}}. \bibinfo{pages}{413--422}.
\newblock


\bibitem[\protect\citeauthoryear{Hopper}{Hopper}{2014}]%
        {hopper2014bringing}
\bibfield{author}{\bibinfo{person}{Susan~B Hopper}.}
  \bibinfo{year}{2014}\natexlab{}.
\newblock \showarticletitle{Bringing the world to the classroom through
  videoconferencing and project-based learning}.
\newblock \bibinfo{journal}{\emph{TechTrends}} \bibinfo{volume}{58},
  \bibinfo{number}{3} (\bibinfo{year}{2014}), \bibinfo{pages}{78--89}.
\newblock


\bibitem[\protect\citeauthoryear{Hrastinski}{Hrastinski}{2008}]%
        {hrastinski2008asynchronous}
\bibfield{author}{\bibinfo{person}{Stefan Hrastinski}.}
  \bibinfo{year}{2008}\natexlab{}.
\newblock \showarticletitle{Asynchronous and synchronous e-learning}.
\newblock \bibinfo{journal}{\emph{Educause quarterly}} \bibinfo{volume}{31},
  \bibinfo{number}{4} (\bibinfo{year}{2008}), \bibinfo{pages}{51--55}.
\newblock


\bibitem[\protect\citeauthoryear{Hussa}{Hussa}{1996}]%
        {hussa1996distance}
\bibfield{author}{\bibinfo{person}{Jukka Hussa}.}
  \bibinfo{year}{1996}\natexlab{}.
\newblock \showarticletitle{Distance education in the school environment:
  Integrating remote classrooms by video conferencing}.
\newblock \bibinfo{journal}{\emph{Journal of Open, Flexible, and Distance
  Learning}} \bibinfo{volume}{2}, \bibinfo{number}{1} (\bibinfo{year}{1996}),
  \bibinfo{pages}{34--44}.
\newblock


\bibitem[\protect\citeauthoryear{Jones, Burke, and Westman}{Jones
  et~al\mbox{.}}{2013}]%
        {jones2013work}
\bibfield{author}{\bibinfo{person}{Fiona Jones}, \bibinfo{person}{Ronald~J
  Burke}, {and} \bibinfo{person}{Mina Westman}.}
  \bibinfo{year}{2013}\natexlab{}.
\newblock \bibinfo{booktitle}{\emph{Work-life balance: A psychological
  perspective}}.
\newblock \bibinfo{publisher}{Psychology Press}.
\newblock


\bibitem[\protect\citeauthoryear{Jones, Moldovan, Raban, and Butler}{Jones
  et~al\mbox{.}}{2008}]%
        {jones2008empirical}
\bibfield{author}{\bibinfo{person}{Quentin Jones}, \bibinfo{person}{Mihai
  Moldovan}, \bibinfo{person}{Daphne Raban}, {and} \bibinfo{person}{Brian
  Butler}.} \bibinfo{year}{2008}\natexlab{}.
\newblock \showarticletitle{Empirical evidence of information overload
  constraining chat channel community interactions}. In
  \bibinfo{booktitle}{\emph{Proceedings of the 2008 ACM conference on Computer
  supported cooperative work}}. \bibinfo{pages}{323--332}.
\newblock


\bibitem[\protect\citeauthoryear{Juhlin, Engstr{\"o}m, and Reponen}{Juhlin
  et~al\mbox{.}}{2010}]%
        {juhlin2010mobile}
\bibfield{author}{\bibinfo{person}{Oskar Juhlin}, \bibinfo{person}{Arvid
  Engstr{\"o}m}, {and} \bibinfo{person}{Erika Reponen}.}
  \bibinfo{year}{2010}\natexlab{}.
\newblock \showarticletitle{Mobile broadcasting: the whats and hows of live
  video as a social medium}. In \bibinfo{booktitle}{\emph{Proceedings of the
  12th international conference on Human computer interaction with mobile
  devices and services}}. \bibinfo{pages}{35--44}.
\newblock


\bibitem[\protect\citeauthoryear{Junuzovic, Inkpen, Hegde, and Zhang}{Junuzovic
  et~al\mbox{.}}{2011}]%
        {junuzovic2011towards}
\bibfield{author}{\bibinfo{person}{Sasa Junuzovic}, \bibinfo{person}{Kori
  Inkpen}, \bibinfo{person}{Rajesh Hegde}, {and} \bibinfo{person}{Zhengyou
  Zhang}.} \bibinfo{year}{2011}\natexlab{}.
\newblock \showarticletitle{Towards ideal window layouts for multi-party,
  gaze-aware desktop videoconferencing}.
\newblock  (\bibinfo{year}{2011}).
\newblock


\bibitem[\protect\citeauthoryear{Kharrufa, Rix, Osadchiy, Preston, and
  Olivier}{Kharrufa et~al\mbox{.}}{2017}]%
        {kharrufa2017group}
\bibfield{author}{\bibinfo{person}{Ahmed Kharrufa}, \bibinfo{person}{Sally
  Rix}, \bibinfo{person}{Timur Osadchiy}, \bibinfo{person}{Anne Preston}, {and}
  \bibinfo{person}{Patrick Olivier}.} \bibinfo{year}{2017}\natexlab{}.
\newblock \showarticletitle{Group Spinner: recognizing and visualizing learning
  in the classroom for reflection, communication, and planning}. In
  \bibinfo{booktitle}{\emph{Proceedings of the 2017 CHI Conference on Human
  Factors in Computing Systems}}. \bibinfo{pages}{5556--5567}.
\newblock


\bibitem[\protect\citeauthoryear{Koehne, Shih, and Olson}{Koehne
  et~al\mbox{.}}{2012}]%
        {koehne2012remote}
\bibfield{author}{\bibinfo{person}{Benjamin Koehne}, \bibinfo{person}{Patrick~C
  Shih}, {and} \bibinfo{person}{Judith~S Olson}.}
  \bibinfo{year}{2012}\natexlab{}.
\newblock \showarticletitle{Remote and alone: coping with being the remote
  member on the team}. In \bibinfo{booktitle}{\emph{Proceedings of the ACM 2012
  conference on Computer Supported Cooperative Work}}.
  \bibinfo{pages}{1257--1266}.
\newblock


\bibitem[\protect\citeauthoryear{Kotturi, Kulkarni, Bernstein, and
  Klemmer}{Kotturi et~al\mbox{.}}{2015}]%
        {kotturi2015structure}
\bibfield{author}{\bibinfo{person}{Yasmine Kotturi}, \bibinfo{person}{Chinmay~E
  Kulkarni}, \bibinfo{person}{Michael~S Bernstein}, {and}
  \bibinfo{person}{Scott Klemmer}.} \bibinfo{year}{2015}\natexlab{}.
\newblock \showarticletitle{Structure and messaging techniques for online peer
  learning systems that increase stickiness}. In
  \bibinfo{booktitle}{\emph{Proceedings of the Second (2015) ACM Conference on
  Learning@ Scale}}. \bibinfo{pages}{31--38}.
\newblock


\bibitem[\protect\citeauthoryear{Kovacs}{Kovacs}{2016}]%
        {kovacs2016effects}
\bibfield{author}{\bibinfo{person}{Geza Kovacs}.}
  \bibinfo{year}{2016}\natexlab{}.
\newblock \showarticletitle{Effects of in-video quizzes on MOOC lecture
  viewing}. In \bibinfo{booktitle}{\emph{Proceedings of the third (2016) ACM
  conference on Learning@ Scale}}. \bibinfo{pages}{31--40}.
\newblock


\bibitem[\protect\citeauthoryear{Kulkarni, Cambre, Kotturi, Bernstein, and
  Klemmer}{Kulkarni et~al\mbox{.}}{2015}]%
        {kulkarni2015talkabout}
\bibfield{author}{\bibinfo{person}{Chinmay Kulkarni}, \bibinfo{person}{Julia
  Cambre}, \bibinfo{person}{Yasmine Kotturi}, \bibinfo{person}{Michael~S
  Bernstein}, {and} \bibinfo{person}{Scott~R Klemmer}.}
  \bibinfo{year}{2015}\natexlab{}.
\newblock \showarticletitle{Talkabout: Making distance matter with small groups
  in massive classes}. In \bibinfo{booktitle}{\emph{Proceedings of the 18th ACM
  Conference on Computer Supported Cooperative Work \& Social Computing}}.
  \bibinfo{pages}{1116--1128}.
\newblock


\bibitem[\protect\citeauthoryear{Lawson, Comber, Gage, and
  Cullum-Hanshaw}{Lawson et~al\mbox{.}}{2010}]%
        {lawson2010images}
\bibfield{author}{\bibinfo{person}{Tony Lawson}, \bibinfo{person}{Chris
  Comber}, \bibinfo{person}{Jenny Gage}, {and} \bibinfo{person}{Adrian
  Cullum-Hanshaw}.} \bibinfo{year}{2010}\natexlab{}.
\newblock \showarticletitle{Images of the future for education?
  Videoconferencing: A literature review}.
\newblock \bibinfo{journal}{\emph{Technology, pedagogy and education}}
  \bibinfo{volume}{19}, \bibinfo{number}{3} (\bibinfo{year}{2010}),
  \bibinfo{pages}{295--314}.
\newblock


\bibitem[\protect\citeauthoryear{Lessel, Vielhauer, and Kr{\"u}ger}{Lessel
  et~al\mbox{.}}{2017}]%
        {lessel2017expanding}
\bibfield{author}{\bibinfo{person}{Pascal Lessel}, \bibinfo{person}{Alexander
  Vielhauer}, {and} \bibinfo{person}{Antonio Kr{\"u}ger}.}
  \bibinfo{year}{2017}\natexlab{}.
\newblock \showarticletitle{Expanding video game live-streams with enhanced
  communication channels: A case study}. In
  \bibinfo{booktitle}{\emph{Proceedings of the 2017 CHI Conference on Human
  Factors in Computing Systems}}. \bibinfo{pages}{1571--1576}.
\newblock


\bibitem[\protect\citeauthoryear{Liu, Zhang, Qiao, Zhou, and Coates}{Liu
  et~al\mbox{.}}{2020}]%
        {liu2020ensuring}
\bibfield{author}{\bibinfo{person}{Yang Liu}, \bibinfo{person}{Yu Zhang},
  \bibinfo{person}{Weifeng Qiao}, \bibinfo{person}{Lu Zhou}, {and}
  \bibinfo{person}{Hamish Coates}.} \bibinfo{year}{2020}\natexlab{}.
\newblock \showarticletitle{Ensuring the sustainability of university learning:
  Case study of a leading Chinese University}.
\newblock \bibinfo{journal}{\emph{Sustainability}} \bibinfo{volume}{12},
  \bibinfo{number}{17} (\bibinfo{year}{2020}), \bibinfo{pages}{6929}.
\newblock


\bibitem[\protect\citeauthoryear{Lu, Annett, Fan, and Wigdor}{Lu
  et~al\mbox{.}}{2019b}]%
        {lu2019feel}
\bibfield{author}{\bibinfo{person}{Zhicong Lu}, \bibinfo{person}{Michelle
  Annett}, \bibinfo{person}{Mingming Fan}, {and} \bibinfo{person}{Daniel
  Wigdor}.} \bibinfo{year}{2019}\natexlab{b}.
\newblock \showarticletitle{" I feel it is my responsibility to stream"
  Streaming and Engaging with Intangible Cultural Heritage through
  Livestreaming}. In \bibinfo{booktitle}{\emph{Proceedings of the 2019 CHI
  Conference on Human Factors in Computing Systems}}. \bibinfo{pages}{1--14}.
\newblock


\bibitem[\protect\citeauthoryear{Lu, Annett, and Wigdor}{Lu
  et~al\mbox{.}}{2019a}]%
        {lu2019vicariously}
\bibfield{author}{\bibinfo{person}{Zhicong Lu}, \bibinfo{person}{Michelle
  Annett}, {and} \bibinfo{person}{Daniel Wigdor}.}
  \bibinfo{year}{2019}\natexlab{a}.
\newblock \showarticletitle{Vicariously experiencing it all without going
  outside: A study of outdoor livestreaming in China}.
\newblock \bibinfo{journal}{\emph{Proceedings of the ACM on Human-Computer
  Interaction}} \bibinfo{volume}{3}, \bibinfo{number}{CSCW}
  (\bibinfo{year}{2019}), \bibinfo{pages}{1--28}.
\newblock


\bibitem[\protect\citeauthoryear{Lu, Heo, and Wigdor}{Lu
  et~al\mbox{.}}{2018a}]%
        {lu2018streamwiki}
\bibfield{author}{\bibinfo{person}{Zhicong Lu}, \bibinfo{person}{Seongkook
  Heo}, {and} \bibinfo{person}{Daniel~J Wigdor}.}
  \bibinfo{year}{2018}\natexlab{a}.
\newblock \showarticletitle{Streamwiki: Enabling viewers of knowledge sharing
  live streams to collaboratively generate archival documentation for effective
  in-stream and post hoc learning}.
\newblock \bibinfo{journal}{\emph{Proceedings of the ACM on Human-Computer
  Interaction}} \bibinfo{volume}{2}, \bibinfo{number}{CSCW}
  (\bibinfo{year}{2018}), \bibinfo{pages}{1--26}.
\newblock


\bibitem[\protect\citeauthoryear{Lu, Xia, Heo, and Wigdor}{Lu
  et~al\mbox{.}}{2018b}]%
        {lu2018you}
\bibfield{author}{\bibinfo{person}{Zhicong Lu}, \bibinfo{person}{Haijun Xia},
  \bibinfo{person}{Seongkook Heo}, {and} \bibinfo{person}{Daniel Wigdor}.}
  \bibinfo{year}{2018}\natexlab{b}.
\newblock \showarticletitle{You watch, you give, and you engage: a study of
  live streaming practices in China}. In \bibinfo{booktitle}{\emph{Proceedings
  of the 2018 CHI conference on human factors in computing systems}}.
  \bibinfo{pages}{1--13}.
\newblock


\bibitem[\protect\citeauthoryear{Ma and Cao}{Ma and Cao}{2017}]%
        {ma2017video}
\bibfield{author}{\bibinfo{person}{Xiaojuan Ma} {and} \bibinfo{person}{Nan
  Cao}.} \bibinfo{year}{2017}\natexlab{}.
\newblock \showarticletitle{Video-based evanescent, anonymous, asynchronous
  social interaction: Motivation and adaption to medium}. In
  \bibinfo{booktitle}{\emph{Proceedings of the 2017 ACM Conference on Computer
  Supported Cooperative Work and Social Computing}}. \bibinfo{pages}{770--782}.
\newblock


\bibitem[\protect\citeauthoryear{Ma, Hancock, and Naaman}{Ma
  et~al\mbox{.}}{2016}]%
        {ma2016anonymity}
\bibfield{author}{\bibinfo{person}{Xiao Ma}, \bibinfo{person}{Jeff Hancock},
  {and} \bibinfo{person}{Mor Naaman}.} \bibinfo{year}{2016}\natexlab{}.
\newblock \showarticletitle{Anonymity, intimacy and self-disclosure in social
  media}. In \bibinfo{booktitle}{\emph{Proceedings of the 2016 CHI conference
  on human factors in computing systems}}. \bibinfo{pages}{3857--3869}.
\newblock


\bibitem[\protect\citeauthoryear{Macaranas, Venolia, Inkpen, and
  Tang}{Macaranas et~al\mbox{.}}{2013}]%
        {macaranas2013sharing}
\bibfield{author}{\bibinfo{person}{Anna Macaranas}, \bibinfo{person}{Gina
  Venolia}, \bibinfo{person}{Kori Inkpen}, {and} \bibinfo{person}{John Tang}.}
  \bibinfo{year}{2013}\natexlab{}.
\newblock \showarticletitle{Sharing Experiences over Video: watching video
  programs together at a distance}. In \bibinfo{booktitle}{\emph{IFIP
  Conference on Human-Computer Interaction}}. Springer,
  \bibinfo{pages}{73--90}.
\newblock


\bibitem[\protect\citeauthoryear{McMillan}{McMillan}{1996}]%
        {mcmillan1996sense}
\bibfield{author}{\bibinfo{person}{David~W McMillan}.}
  \bibinfo{year}{1996}\natexlab{}.
\newblock \showarticletitle{Sense of community}.
\newblock \bibinfo{journal}{\emph{Journal of community psychology}}
  \bibinfo{volume}{24}, \bibinfo{number}{4} (\bibinfo{year}{1996}),
  \bibinfo{pages}{315--325}.
\newblock


\bibitem[\protect\citeauthoryear{McMillan and Chavis}{McMillan and
  Chavis}{1986}]%
        {mcmillan1986sense}
\bibfield{author}{\bibinfo{person}{David~W McMillan} {and}
  \bibinfo{person}{David~M Chavis}.} \bibinfo{year}{1986}\natexlab{}.
\newblock \showarticletitle{Sense of community: A definition and theory}.
\newblock \bibinfo{journal}{\emph{Journal of community psychology}}
  \bibinfo{volume}{14}, \bibinfo{number}{1} (\bibinfo{year}{1986}),
  \bibinfo{pages}{6--23}.
\newblock


\bibitem[\protect\citeauthoryear{Mehrotra, Chen, Zhang, and Chou}{Mehrotra
  et~al\mbox{.}}{2011}]%
        {mehrotra2011realistic}
\bibfield{author}{\bibinfo{person}{Sanjeev Mehrotra}, \bibinfo{person}{Wei-ge
  Chen}, \bibinfo{person}{Zhengyou Zhang}, {and} \bibinfo{person}{Philip~A
  Chou}.} \bibinfo{year}{2011}\natexlab{}.
\newblock \showarticletitle{Realistic audio in immersive video conferencing}.
  In \bibinfo{booktitle}{\emph{2011 IEEE International Conference on Multimedia
  and Expo}}. IEEE, \bibinfo{pages}{1--4}.
\newblock


\bibitem[\protect\citeauthoryear{Mitchell, Agle, and Wood}{Mitchell
  et~al\mbox{.}}{1997}]%
        {mitchell1997toward}
\bibfield{author}{\bibinfo{person}{Ronald~K Mitchell},
  \bibinfo{person}{Bradley~R Agle}, {and} \bibinfo{person}{Donna~J Wood}.}
  \bibinfo{year}{1997}\natexlab{}.
\newblock \showarticletitle{Toward a theory of stakeholder identification and
  salience: Defining the principle of who and what really counts}.
\newblock \bibinfo{journal}{\emph{Academy of management review}}
  \bibinfo{volume}{22}, \bibinfo{number}{4} (\bibinfo{year}{1997}),
  \bibinfo{pages}{853--886}.
\newblock


\bibitem[\protect\citeauthoryear{Moeller and Seehuus}{Moeller and
  Seehuus}{2019}]%
        {moeller2019loneliness}
\bibfield{author}{\bibinfo{person}{Robert~W Moeller} {and}
  \bibinfo{person}{Martin Seehuus}.} \bibinfo{year}{2019}\natexlab{}.
\newblock \showarticletitle{Loneliness as a mediator for college students'
  social skills and experiences of depression and anxiety}.
\newblock \bibinfo{journal}{\emph{Journal of adolescence}}
  \bibinfo{volume}{73} (\bibinfo{year}{2019}), \bibinfo{pages}{1--13}.
\newblock


\bibitem[\protect\citeauthoryear{Newhart and Olson}{Newhart and Olson}{2017}]%
        {newhart2017my}
\bibfield{author}{\bibinfo{person}{Veronica~Ahumada Newhart} {and}
  \bibinfo{person}{Judith~S Olson}.} \bibinfo{year}{2017}\natexlab{}.
\newblock \showarticletitle{My student is a robot: How schools manage
  telepresence experiences for students}. In
  \bibinfo{booktitle}{\emph{Proceedings of the 2017 CHI conference on human
  factors in computing systems}}. \bibinfo{pages}{342--347}.
\newblock


\bibitem[\protect\citeauthoryear{Olson and Olson}{Olson and Olson}{2000}]%
        {olson2000distance}
\bibfield{author}{\bibinfo{person}{Gary~M Olson} {and}
  \bibinfo{person}{Judith~S Olson}.} \bibinfo{year}{2000}\natexlab{}.
\newblock \showarticletitle{Distance matters}.
\newblock \bibinfo{journal}{\emph{Human--computer interaction}}
  \bibinfo{volume}{15}, \bibinfo{number}{2-3} (\bibinfo{year}{2000}),
  \bibinfo{pages}{139--178}.
\newblock


\bibitem[\protect\citeauthoryear{Poon, Giroux, Eloundou-Enyegue,
  Guimbreti{\`e}re, and Dell}{Poon et~al\mbox{.}}{2019}]%
        {poon2019engaging}
\bibfield{author}{\bibinfo{person}{Anthony Poon}, \bibinfo{person}{Sarah
  Giroux}, \bibinfo{person}{Parfait Eloundou-Enyegue},
  \bibinfo{person}{Fran{\c{c}}ois Guimbreti{\`e}re}, {and}
  \bibinfo{person}{Nicola Dell}.} \bibinfo{year}{2019}\natexlab{}.
\newblock \showarticletitle{Engaging high school students in cameroon with exam
  practice quizzes via sms and whatsapp}. In
  \bibinfo{booktitle}{\emph{Proceedings of the 2019 CHI Conference on Human
  Factors in Computing Systems}}. \bibinfo{pages}{1--13}.
\newblock


\bibitem[\protect\citeauthoryear{Riggio, Watring, and Throckmorton}{Riggio
  et~al\mbox{.}}{1993}]%
        {riggio1993social}
\bibfield{author}{\bibinfo{person}{Ronald~E Riggio}, \bibinfo{person}{Kristin~P
  Watring}, {and} \bibinfo{person}{Barbara Throckmorton}.}
  \bibinfo{year}{1993}\natexlab{}.
\newblock \showarticletitle{Social skills, social support, and psychosocial
  adjustment}.
\newblock \bibinfo{journal}{\emph{Personality and Individual Differences}}
  \bibinfo{volume}{15}, \bibinfo{number}{3} (\bibinfo{year}{1993}),
  \bibinfo{pages}{275--280}.
\newblock


\bibitem[\protect\citeauthoryear{Sawyer}{Sawyer}{2005}]%
        {sawyer2005cambridge}
\bibfield{author}{\bibinfo{person}{R~Keith Sawyer}.}
  \bibinfo{year}{2005}\natexlab{}.
\newblock \bibinfo{booktitle}{\emph{The Cambridge handbook of the learning
  sciences}}.
\newblock \bibinfo{publisher}{Cambridge University Press}.
\newblock


\bibitem[\protect\citeauthoryear{Senior, Pankanti, Hampapur, Brown, Tian, Ekin,
  Connell, Shu, and Lu}{Senior et~al\mbox{.}}{2005}]%
        {senior2005enabling}
\bibfield{author}{\bibinfo{person}{Andrew Senior}, \bibinfo{person}{Sharath
  Pankanti}, \bibinfo{person}{Arun Hampapur}, \bibinfo{person}{Lisa Brown},
  \bibinfo{person}{Ying-Li Tian}, \bibinfo{person}{Ahmet Ekin},
  \bibinfo{person}{Jonathan Connell}, \bibinfo{person}{Chiao~Fe Shu}, {and}
  \bibinfo{person}{Max Lu}.} \bibinfo{year}{2005}\natexlab{}.
\newblock \showarticletitle{Enabling video privacy through computer vision}.
\newblock \bibinfo{journal}{\emph{IEEE Security \& Privacy}}
  \bibinfo{volume}{3}, \bibinfo{number}{3} (\bibinfo{year}{2005}),
  \bibinfo{pages}{50--57}.
\newblock


\bibitem[\protect\citeauthoryear{Shannon, Sciuto, Hu, Dow, and Hammer}{Shannon
  et~al\mbox{.}}{2017}]%
        {shannon2017better}
\bibfield{author}{\bibinfo{person}{Amy Shannon}, \bibinfo{person}{Alex Sciuto},
  \bibinfo{person}{Danielle Hu}, \bibinfo{person}{Steven~P Dow}, {and}
  \bibinfo{person}{Jessica Hammer}.} \bibinfo{year}{2017}\natexlab{}.
\newblock \showarticletitle{Better Organization or a Source of Distraction?
  Introducing Digital Peer Feedback to a Paper-Based Classroom}. In
  \bibinfo{booktitle}{\emph{Proceedings of the 2017 CHI Conference on Human
  Factors in Computing Systems}}. \bibinfo{pages}{5545--5555}.
\newblock


\bibitem[\protect\citeauthoryear{Sun, Rosson, and Carroll}{Sun
  et~al\mbox{.}}{2018}]%
        {sun2018community}
\bibfield{author}{\bibinfo{person}{Na Sun}, \bibinfo{person}{Mary~Beth Rosson},
  {and} \bibinfo{person}{John~M Carroll}.} \bibinfo{year}{2018}\natexlab{}.
\newblock \showarticletitle{Where is community among online learners? Identity,
  efficacy and personal ties}. In \bibinfo{booktitle}{\emph{Proceedings of the
  2018 chi conference on human factors in computing systems}}.
  \bibinfo{pages}{1--13}.
\newblock


\bibitem[\protect\citeauthoryear{Sun, Wang, and Rosson}{Sun
  et~al\mbox{.}}{2019b}]%
        {sun2019distance}
\bibfield{author}{\bibinfo{person}{Na Sun}, \bibinfo{person}{Xiying Wang},
  {and} \bibinfo{person}{Mary~Beth Rosson}.} \bibinfo{year}{2019}\natexlab{b}.
\newblock \showarticletitle{How Do Distance Learners Connect?}. In
  \bibinfo{booktitle}{\emph{Proceedings of the 2019 CHI Conference on Human
  Factors in Computing Systems}}. \bibinfo{pages}{1--12}.
\newblock


\bibitem[\protect\citeauthoryear{Sun, Li, Tian, Fan, and Wang}{Sun
  et~al\mbox{.}}{2019a}]%
        {sun2019presenters}
\bibfield{author}{\bibinfo{person}{Wei Sun}, \bibinfo{person}{Yunzhi Li},
  \bibinfo{person}{Feng Tian}, \bibinfo{person}{Xiangmin Fan}, {and}
  \bibinfo{person}{Hongan Wang}.} \bibinfo{year}{2019}\natexlab{a}.
\newblock \showarticletitle{How Presenters Perceive and React to Audience Flow
  Prediction In-situ: An Explorative Study of Live Online Lectures}.
\newblock \bibinfo{journal}{\emph{Proceedings of the ACM on Human-Computer
  Interaction}} \bibinfo{volume}{3}, \bibinfo{number}{CSCW}
  (\bibinfo{year}{2019}), \bibinfo{pages}{1--19}.
\newblock


\bibitem[\protect\citeauthoryear{Tang, Venolia, and Inkpen}{Tang
  et~al\mbox{.}}{2016}]%
        {tang2016meerkat}
\bibfield{author}{\bibinfo{person}{John~C Tang}, \bibinfo{person}{Gina
  Venolia}, {and} \bibinfo{person}{Kori~M Inkpen}.}
  \bibinfo{year}{2016}\natexlab{}.
\newblock \showarticletitle{Meerkat and periscope: I stream, you stream, apps
  stream for live streams}. In \bibinfo{booktitle}{\emph{Proceedings of the
  2016 CHI Conference on Human Factors in Computing Systems}}.
  \bibinfo{pages}{4770--4780}.
\newblock


\bibitem[\protect\citeauthoryear{Weinstein}{Weinstein}{1987}]%
        {weinstein1987fostering}
\bibfield{author}{\bibinfo{person}{Claire~E Weinstein}.}
  \bibinfo{year}{1987}\natexlab{}.
\newblock \showarticletitle{Fostering learning autonomy through the use of
  learning strategies}.
\newblock \bibinfo{journal}{\emph{Journal of reading}} \bibinfo{volume}{30},
  \bibinfo{number}{7} (\bibinfo{year}{1987}), \bibinfo{pages}{590--595}.
\newblock


\bibitem[\protect\citeauthoryear{Wen, Maki, Dow, Herbsleb, and Rose}{Wen
  et~al\mbox{.}}{2017}]%
        {wen2017supporting}
\bibfield{author}{\bibinfo{person}{Miaomiao Wen}, \bibinfo{person}{Keith Maki},
  \bibinfo{person}{Steven Dow}, \bibinfo{person}{James~D Herbsleb}, {and}
  \bibinfo{person}{Carolyn Rose}.} \bibinfo{year}{2017}\natexlab{}.
\newblock \showarticletitle{Supporting virtual team formation through
  community-wide deliberation}.
\newblock \bibinfo{journal}{\emph{Proceedings of the ACM on Human-Computer
  Interaction}} \bibinfo{volume}{1}, \bibinfo{number}{CSCW}
  (\bibinfo{year}{2017}), \bibinfo{pages}{1--19}.
\newblock


\bibitem[\protect\citeauthoryear{Whiting, Blaising, Barreau, Fiuza, Marda,
  Valentine, and Bernstein}{Whiting et~al\mbox{.}}{2019}]%
        {whiting2019did}
\bibfield{author}{\bibinfo{person}{Mark~E Whiting}, \bibinfo{person}{Allie
  Blaising}, \bibinfo{person}{Chloe Barreau}, \bibinfo{person}{Laura Fiuza},
  \bibinfo{person}{Nik Marda}, \bibinfo{person}{Melissa Valentine}, {and}
  \bibinfo{person}{Michael~S Bernstein}.} \bibinfo{year}{2019}\natexlab{}.
\newblock \showarticletitle{Did It Have To End This Way? Understanding the
  Consistency of Team Fracture}.
\newblock \bibinfo{journal}{\emph{Proceedings of the ACM on Human-Computer
  Interaction}} \bibinfo{volume}{3}, \bibinfo{number}{CSCW}
  (\bibinfo{year}{2019}), \bibinfo{pages}{1--23}.
\newblock


\bibitem[\protect\citeauthoryear{Whiting, Gao, Xing, Diarrassouba, Nguyen, and
  Bernstein}{Whiting et~al\mbox{.}}{2020}]%
        {whiting2020parallel}
\bibfield{author}{\bibinfo{person}{Mark~E Whiting}, \bibinfo{person}{Irena
  Gao}, \bibinfo{person}{Michelle Xing}, \bibinfo{person}{N'godjigui~Junior
  Diarrassouba}, \bibinfo{person}{Tonya Nguyen}, {and}
  \bibinfo{person}{Michael~S Bernstein}.} \bibinfo{year}{2020}\natexlab{}.
\newblock \showarticletitle{Parallel Worlds: Repeated Initializations of the
  Same Team To Improve Team Viability}.
\newblock \bibinfo{journal}{\emph{Proceedings of the ACM on Human-Computer
  Interaction}} \bibinfo{volume}{4}, \bibinfo{number}{CSCW1}
  (\bibinfo{year}{2020}), \bibinfo{pages}{1--22}.
\newblock


\bibitem[\protect\citeauthoryear{Xiao and Wang}{Xiao and Wang}{2017}]%
        {xiao2017undertanding}
\bibfield{author}{\bibinfo{person}{Xiang Xiao} {and} \bibinfo{person}{Jingtao
  Wang}.} \bibinfo{year}{2017}\natexlab{}.
\newblock \showarticletitle{Undertanding and detecting divided attention in
  mobile mooc learning}. In \bibinfo{booktitle}{\emph{Proceedings of the 2017
  CHI Conference on Human Factors in Computing Systems}}.
  \bibinfo{pages}{2411--2415}.
\newblock


\bibitem[\protect\citeauthoryear{Yang and Huang}{Yang and Huang}{2020}]%
        {yang2020turn}
\bibfield{author}{\bibinfo{person}{Bin Yang} {and} \bibinfo{person}{Cheng
  Huang}.} \bibinfo{year}{2020}\natexlab{}.
\newblock \showarticletitle{Turn crisis into opportunity in response to
  COVID-19: experience from a Chinese University and future prospects}.
\newblock \bibinfo{journal}{\emph{Studies in Higher Education}}
  (\bibinfo{year}{2020}), \bibinfo{pages}{1--12}.
\newblock


\bibitem[\protect\citeauthoryear{Yang, Lee, Shin, and Kim}{Yang
  et~al\mbox{.}}{2020}]%
        {yang2020snapstream}
\bibfield{author}{\bibinfo{person}{Saelyne Yang}, \bibinfo{person}{Changyoon
  Lee}, \bibinfo{person}{Hijung~Valentina Shin}, {and} \bibinfo{person}{Juho
  Kim}.} \bibinfo{year}{2020}\natexlab{}.
\newblock \showarticletitle{Snapstream: Snapshot-based Interaction in Live
  Streaming for Visual Art}. In \bibinfo{booktitle}{\emph{Proceedings of the
  2020 CHI Conference on Human Factors in Computing Systems}}.
  \bibinfo{pages}{1--12}.
\newblock


\bibitem[\protect\citeauthoryear{Zheng, Rosson, Shih, and Carroll}{Zheng
  et~al\mbox{.}}{2015}]%
        {zheng2015understanding}
\bibfield{author}{\bibinfo{person}{Saijing Zheng}, \bibinfo{person}{Mary~Beth
  Rosson}, \bibinfo{person}{Patrick~C Shih}, {and} \bibinfo{person}{John~M
  Carroll}.} \bibinfo{year}{2015}\natexlab{}.
\newblock \showarticletitle{Understanding student motivation, behaviors and
  perceptions in MOOCs}. In \bibinfo{booktitle}{\emph{Proceedings of the 18th
  ACM conference on computer supported cooperative work \& social computing}}.
  \bibinfo{pages}{1882--1895}.
\newblock


\bibitem[\protect\citeauthoryear{Zheng, Wisniewski, Rosson, and Carroll}{Zheng
  et~al\mbox{.}}{2016}]%
        {zheng2016ask}
\bibfield{author}{\bibinfo{person}{Saijing Zheng}, \bibinfo{person}{Pamela
  Wisniewski}, \bibinfo{person}{Mary~Beth Rosson}, {and}
  \bibinfo{person}{John~M Carroll}.} \bibinfo{year}{2016}\natexlab{}.
\newblock \showarticletitle{Ask the instructors: Motivations and challenges of
  teaching massive open online courses}. In
  \bibinfo{booktitle}{\emph{Proceedings of the 19th ACM Conference on
  Computer-Supported Cooperative Work \& Social Computing}}.
  \bibinfo{pages}{206--221}.
\newblock


\end{thebibliography}


\end{document}